\documentclass[preprint,notoc]{JHEP3}
\usepackage{amssymb,amsmath,bbm,bm,dsfont}
\usepackage{epsfig}
\DeclareOldFontCommand{\rm}{\normalfont\rmfamily}{\mathrm}
\DeclareOldFontCommand{\bf}{\normalfont\bfseries}{\mathbf}
\def\tz{\widetilde Z}

\def\ttau{\tilde \tau}
\def\tnu{\tilde\nu}
\def\ttop{\tilde t}
\def\te{\tilde e}

\def\tu{\tilde u}

\def\td{\tilde d}
\def\tQ{\tilde Q}
\def\tL{\tilde L}
\def\tH{\tilde H}
\def\th{\tilde h}
\def\tg{\tilde g}
\def\tB{\tilde B}
\def\tW{\tilde W}

\def\mns{{\bf m}^2_{\tnu_R}}

\def\bfu{{\bf f}_u}
\def\bfd{{\bf f}_d}
\def\bfe{{\bf f}_e}
\def\bfn{{\bf f}_{\nu}}
\def\banu{{\bf a}_{\nu}}
\def\bbnu{{\bf b}_{\nu}}
\def\byu{{\bf f}_u^T}
\def\byd{{\bf f}_d^T}
\def\bye{{\bf f}_e^T}

\def\MSb{\overline {MS}}
\def\DRb{\overline {DR}}
\def\bcm2{\bm{\mathcal{M}}^2}
\newcommand{\beq}{\begin{equation}}
\newcommand{\eeq}{\end{equation}}
\def\bea{\begin{eqnarray}}  
\def\eea{\end{eqnarray}}  
\DeclareMathOperator{\Tr}{Tr}

\title{SUSY dark matter and lepton flavor violation}
\author{Vernon Barger$^a$, Danny Marfatia$^b$, Azar Mustafayev$^b$ and Ali Soleimani$^b$\\
        $^a$Department of Physics, University of Wisconsin, Madison, WI 53706, USA\\
        $^b$Department of Physics \& Astronomy, University of Kansas, Lawrence, KS 66045, USA\\
E-mail: \email{barger@physics.wisc.edu}, \email{marfatia@ku.edu}, \email{amustaf@ku.edu},
        \email{alis@ku.edu}}

\abstract{
We study lepton flavor-violating (LFV) processes within a supersymmetric type-I seesaw framework with 
flavor-blind universal boundary conditions,
properly accounting for the effect of the neutrino sector on the dark matter relic abundance.
We consider several possibilities for the neutrino Yukawa coupling matrix
and show that in regions of SUSY parameter space that yield the correct neutralino relic density,  
LFV rates can differ from naive estimates by up to two orders of magnitude. Contrary to common belief,
we find that current LFV limits do not exclude neutrino Yukawa couplings larger than top Yukawa couplings.
We introduce the ISAJET-M program that was used for the computations.
}

\begin{document}

\section{Introduction}

It is now a firmly established experimental fact that neutrinos are massive and mix~\cite{bmw}.
A global fit to current neutrino oscillation data~\cite{Schwetz:2008er} gives the following $3\sigma$ ranges/limits for the mixing parameters 
(in the ``standard parametrization'', Eq.~\ref{eq:mns}),
\begin{align}
\label{nuoscil}
\sin^2 \theta_{12}& =0.304^{+0.066}_{-0.054}\, , & 
   \Delta m_{21}^2 & =(7.65^{+0.69}_{-0.60})\times 10^{-5}\textrm{eV}^2, \nonumber \\
\sin^2 \theta_{23}& =0.50^{+0.17}_{-0.14}\, , & 
   |\Delta m_{31}^2| & =(2.40^{+0.35}_{-0.33})\times 10^{-3}\textrm{eV}^2, \\
\sin^2 \theta_{13}& < 0.056\,. \nonumber
\end{align}
The phases are not constrained by current data. 

The most elegant and popular explanation for small neutrino masses is offered by the seesaw
mechanism~\cite{seesaw1}. Here the Standard Model (SM) is augmented by three right-handed
neutrinos (RHNs). The RHNs transform as singlets under the SM gauge group and thus can have
a Majorana mass, which is taken to be very large, $M_{Maj}\sim 10^{15}$~GeV. One of the
implications of neutrino mixing is the possibility for charged lepton flavor-violating
processes. In the SM-seesaw these rates are highly suppressed by the very large scale
$M_{Maj}$~\cite{lfv_sm}, so that LFV processes are a good probe of new physics.

Supersymmetry (SUSY) is a well motivated possibility for new physics~\cite{Baer:book,susy_books} because, 
among other things, it stabilizes the seesaw
mechanism~\cite{seesaw_stab}
and ameliorates the hierarchy problem appearing from the presence of the very high scale
$M_{Maj}$~\cite{RHN_hier}.
SUSY must be a broken symmetry that is parametrized by soft SUSY breaking
(SSB) terms in the Lagrangian~\cite{Girardello:1981wz}. In general these terms can have
arbitrary flavor structures that would induce unacceptably large flavor violating effects --
this is the well-known SUSY flavor problem. The simplest solution is to assume a flavor-blind 
SUSY breaking mediation mechanism that will generate flavor universal SSB terms at some
high scale. However, this does not mean that SSB terms will also remain flavor-universal at the weak
scale: flavor-violating terms will be generated by the Yukawa couplings during the evolution from
the high to the weak scale. As a consequence, SUSY contributions to LFV processes are suppressed
by the characteristic mass scale of SUSY particles $M_{SUSY}\sim 1$~TeV (instead of $M_{Maj}$) and thus LFV can be observable.
Many authors have studied these processes under various SUSY model assumptions and seesaw parameters
(see {\it e.g.},
\cite{casas,Masiero:2002jn,hisano,Hisano:1998fj,Petcov:2003zb,hpenguin,arganda,lfv_th13,Ibarra:2008uv,lvfpapers}).

The prediction of LFV rates requires knowledge of the neutrino Yukawa coupling matrix.
However, experimentally measured light neutrino masses and mixings 
do not provide sufficient information to determine it~\cite{casas}. A top-down approach is frequently
adopted in which the neutrino Yukawa matrix is set by a specific SUSY-Grand Unified Theory (GUT), with an $SO(10)$ gauge group being the favorite choice. 
In this work we take a general approach, and consider two cases of Yukawa unification parameters (defined in Eq.~\ref{yukpar}) inspired
by $SO(10)$ relations. For both of these cases we study two scenarios with small and large mixings in the neutrino
Yukawa matrix. 
Our analysis is based on the type-I seesaw mechanism~\cite{seesaw1}.

The existence of a massive, electrically and color neutral, stable weakly interacting particle
that can serve as a cold Dark Matter (DM) candidate is perhaps the most compelling feature of the $R-$parity
conserving Minimal Supersymmetric Standard Model (MSSM). In most cases the DM particle is the lightest neutralino, $\tz_1$~\cite{griest}.
The mass density of DM has been precisely determined by cosmological measurements: a combination of
WMAP CMB data with the baryon acoustic oscillations in galaxy power spectra gives~\cite{wmap}
\bea
\Omega_{DM} h^2 = 0.1120^{+0.0074}_{-0.0076} \qquad (2\sigma)\;,
\label{eq:wmap}
\eea
where $\Omega\equiv \rho/\rho_c$ with $\rho_c$ the critical mass density of the
Universe, and $h$ is the scaled Hubble parameter.
Such a precise determination places severe constraints on new physics scenarios. 
In the simplest SUSY model with universal SSB values at the high scale, mSUGRA
(or CMSSM)~\cite{msugra}, only a few regions of parameter space survive: 
the bulk region~\cite{bulk,Baer:1995nc}, 
the stau~\cite{dm:stau,isared} or stop~\cite{dm:stop}
coannihilation regions, the hyperbolic branch/focus point (HB/FP) region~\cite{dm:fp}, and
the $A$ or $h$ resonance annihilation (Higgs funnel) regions~\cite{Baer:1995nc,Afunnel,hfunnel}.
The benchmark values of mSUGRA input parameters for these regions are listed in Table~\ref{tab:msugra}.

\TABLE{
\begin{tabular}{c||cccc|c}
\hline
Point & $m_0$ & $m_{1/2}$ & $A_0$ & $\tan\beta$ & Region \\
\hline
A &  80 & 170 & -250 & 10 & bulk \\
B & 100 & 500 & 0    & 10 & $\ttau$-coan.\\
C & 150 & 300 & -1095& 5  & $\ttop$-coan.\\
D & 500 & 450 & 0    & 51 & $A$-funnel \\
E & 1370& 300 & 0    & 10 & HB/FP \\
F & 3143& 1000& 0    & 10 & HB/FP \\
G & 2000& 130 & -2000& 10 & $h$-funnel \\
\hline
\end{tabular}
\caption{
Input parameters for benchmark points and corresponding DM-allowed regions of mSUGRA. 
The dimensionful parameters $m_0,\ m_{1/2}$ and $A_0$ are in GeV.  
For all points $\mu >0$ and $m_t = 171$~GeV.}
\label{tab:msugra}}

The introduction of the RHNs and their associated Yukawa couplings changes predictions for some
sparticle masses via new contributions to the renormalization group equations (RGEs). These imprints can be used to
extract neutrino Yukawa couplings in collider experiments~\cite{Baer:2000hx}. 
In previous work~\cite{msugrarhn}, some of us demonstrated that these neutrino-induced changes
in the SUSY mass spectrum can significantly alter the DM (co)annihilation mechanisms 
with concomitant changes in $\Omega_{\tz_1}h^2$ and 
DM direct and indirect detection rates.

The aim of the present work is to study predictions for LFV rates while correctly taking into account the
aforementioned effect on the neutralino relic density. We take a model-independent approach and only consider effects
from RGE running below the unification scale $M_{GUT}$. These two important points distinguish our work 
(which is closer in spirit to the study of Ref.~\cite{Masiero:2002jn})
from the study of Ref.~\cite{Ciuchini:2007ha}. 
We find that proper consideration of the interplay between the neutrino and SUSY sectors can change the predictions for
the LFV rates in WMAP-allowed regions by a factor between $2-100$ compared with naive estimates, 
and that contrary to common
belief, large neutrino Yukawa couplings are not ruled out by current LFV bounds. 

The rest of the paper is organized as follows. In the next section we briefly review LFV processes in SUSY-seesaw
framework and motivate our ansatz for the neutrino Yukawa matrix. Results from numerical analyses are presented in
Section~\ref{sec:results}. Section~\ref{sec:disc} is devoted to the discussion of our findings.
Finally, we present our conclusions in Section~\ref{sec:conclus}.  
Our notation, a description of our code, and Yukawa RGEs with thresholds can be found in the appendices.

\section{SUSY-seesaw and LFV processes}

We begin with a brief discussion of our formalism. Details about our notation and conventions are
relegated to Appendix~\ref{app:notation}.
The superpotential for the MSSM augmented by singlet right-handed neutrinos $\hat{N}^c_i$ is
\beq
\hat{f}=\hat{f}_{MSSM}+
(\bfn)_{ij}\epsilon_{ab}\hat{L}^a_i\hat{H}_u^b\hat{N}^c_j +
\frac{1}{2}({\bf M}_N)_{ij}\hat{N}^c_i \hat{N}^c_j \, ,
\label{eq:RHNspot}
\eeq
where $\hat{f}_{MSSM}$ is the MSSM superpotential shown in Eq.~(\ref{eq:superpot}), 
$\hat{L}$ and $\hat{H}_u$ are, respectively, lepton doublet and
up-higgs superfields, and ${\bf M}_N$ represents the Majorana mass matrix for the (heavy) right-handed neutrinos.
Above the scale $M_{Maj}$ the light neutrino mass matrix is given by the well-known type-I seesaw formula~\cite{seesaw1},
\beq
\mathcal{M}_{\nu} =-\bfn {\bf M}_N^{-1} \bfn^T v^2_u \, ,
\label{eq:seesaw}
\eeq
where $v_u$ is the vacuum expectation value (VEV) of the neutral component $h^0_u$ of the up-type Higgs
doublet $H_u$.
At low energies the RHNs decouple from the theory and the light neutrino mass matrix is 
$\mathcal{M}_{\nu}=-\kappa v^2_u$. Here $\kappa$ is a coupling matrix of the dimension-5 effective operator 
generated by RHNs (see Eq.~(\ref{eq:effspot}) for the definition), which is determined by 
matching conditions (\ref{eq:Kmatch}) at the RHN decoupling thresholds. 

\TABLE{
\begin{tabular}{l|c|c}
\hline
  & Present & Future  \\
\hline
BR($\mu \rightarrow e\gamma$)    & $1.2 \times 10^{-11}$~\cite{mega} & $10^{-13}$~\cite{meg}\\
BR($\tau \rightarrow \mu\gamma $)& $4.5 \times 10^{-8}$~\cite{belle} & $10^{-9}$~\cite{Bona:2007qt} \\
BR($\tau \rightarrow e\gamma$)   & $3.3 \times 10^{-8}$~\cite{babar} & $10^{-9}$~\cite{Bona:2007qt} \\
BR($\mu \rightarrow eee$)        & $1.0 \times 10^{-12}$~\cite{sindrum} & $10^{-14}$~\cite{Marciano:2008zz}\\
BR($\tau \rightarrow \mu\mu\mu $)& $3.2 \times 10^{-8}$~\cite{belle_3l} & $10^{-9}$~\cite{Bona:2007qt} \\
BR($\tau \rightarrow eee$)       & $3.6 \times 10^{-8}$~\cite{belle_3l} & $10^{-9}$~\cite{Bona:2007qt} \\
CR($\mu\,\textrm{Ti}\rightarrow e\,\textrm{Ti}$) & $4.3 \times 10^{-12}$~\cite{sindrum2} &
$10^{-18}$~\cite{prime}\\
CR($\mu\,\textrm{Al}\rightarrow e\,\textrm{Al}$) & - & $10^{-16}$~\cite{mu2e} \\
\hline
\end{tabular}
\caption{Present bounds and projected sensitivities for LFV processes.}
\label{tab:lfv}}

A common solution to the SUSY flavor problem is to assume that at some high scale
sfermion SSB mass-squared matrices are
diagonal and universal in flavor, and the trilinear couplings are proportional to the Yukawa
couplings.
In the framework of mSUGRA extended by right-handed neutrinos (mSUGRA-seesaw) the SSB boundary
conditions at Grand Unification Scale $M_{GUT}$ take the particularly simple form:
\beq
\label{eq:univbc}
{\bf m^2_{Q,U,D,L,E,\tnu_R}} = m_0^2 \mathds{1}\,,\ \ \ \ \ \ 
m^2_{H_u}=m^2_{H_d} = m_0^2\,,\ \ \ \ \ \
{\bf a}_{u,d,e,\nu} = -A_0 {\bf f}_{u,d,e,\nu}\,.
\eeq
Since the Yukawa couplings $\bfe$ and $\bfn$ cannot be simultaneously diagonalized, non-vanishing
off-diagonal elements will be generated in the left-handed slepton mass matrix ${\bm m^2_L}$ via 
renormalization group evolution. In the leading-logarithmic approximation with universal boundary
conditions (\ref{eq:univbc}) the off-diagonal elements are
\beq
 (\bm m^2_L)_{i \neq j} \simeq -\frac{1}{8\pi^2}(3m_0^2+A_0^2)
                          \sum_k (\bfn^T)_{ik}(\bfn^*)_{kj}\log\frac{M_{GUT}}{M_{N_k}}\,,
\label{eq:llog}
\eeq
where $M_{N_k}$ are RHN decoupling scales that are approximately equal to the eigenvalues of the Majorana
mass matrix ${\bf M}_N$. At low energies, these off-diagonal terms induce LFV processes such as
$l_i \rightarrow l_j \gamma$, $l_i \rightarrow 3l_j$ and $l_i \rightarrow l_j$ conversion in nuclei.
Current bounds on LFV processes as well as the projected sensitivities of future experiments are summarized
in Table~\ref{tab:lfv}.

The branching ratio for the flavor-violating radiative decay of a charged lepton is given by
\beq
BR(l_i \rightarrow l_j \gamma)=\frac{\alpha}{4\Gamma(l_i)}m_{l_i}^5(|A_L|^2+|A_R|^2)\,.
\label{eq:llg_full}
\eeq
Here $\alpha$ is the electromagnetic fine structure constant, 
$\Gamma(l_i)$ is the total decay width of the initial lepton, and 
$A_{L,R}$ are form factors for left and right chiralities of the incoming lepton whose full 
expressions in SUSY were obtained in Ref.~\cite{hisano}.
Because $m_{l_i} \gg m_{l_j}$ one has $A_R \gg A_L$ in the case of initial universality such as
(\ref{eq:univbc})~\cite{casas,Hisano:1998fj}. 
In the mass-insertion approximation, the branching ratio can be related to the corresponding off-diagonal element of
the left-handed slepton mass matrix~\cite{Hisano:1998fj},
\beq
BR(l_i \rightarrow l_j \gamma) \simeq 
 BR(l_i \rightarrow l_j \bar{\nu}_j \nu_i)\frac{\alpha^3}{G_F^2 m_s^8}\left|(\bm m^2_L)_{i\neq j}\right|^2 \tan^2 \beta\,, 
\label{eq:llg_mi}
\eeq
where $G_F$ is the Fermi coupling constant and $m_s$ is the characteristic mass scale of the SUSY particles in
the loop. 
In the case of universal boundary conditions (\ref{eq:univbc}), this expression used in conjunction with
the leading-log result (\ref{eq:llog}) well approximates the full expression (\ref{eq:llg_full}),
if one sets~\cite{Petcov:2003zb}
\beq
m_s^8 \simeq 0.5 m_0^2 m_{1/2}^2 \left(m_0^2 + 0.6 m_{1/2}^2\right) ^2\,.
\label{eq:ms}
\eeq

LFV $l_i \rightarrow 3l_j$ decays and $l_i \rightarrow l_j$ conversion occur via $\gamma$-, $Z-$
and Higgs-penguins as well as squark/slepton box diagrams~\cite{hisano}. 
Higgs-penguins dominate in the regime of large $\tan \beta \simeq 60$ and light $H/A$ Higgs boson mass
$\sim 100$~GeV, and enhance rates by up to a few orders of magnitude~\cite{hpenguin}.
However, the latter condition cannot be realized in the universal scenario (\ref{eq:univbc})~\cite{Belyaev:2007xa}. 
It was shown in Ref.~\cite{arganda} that these LFV rates are well described in the universal scanario 
by the same $\gamma$-penguins that contribute to the radiative decays.
The branching ratio for trilepton decays is approximately given by  
\beq
BR(l_i \rightarrow 3l_j) \simeq \frac{\alpha}{3\pi}
     \left( \log \frac{m_{l_i}^2}{m_{l_j}^2}-\frac{11}{4} \right)
     BR(l_i \rightarrow l_j \gamma)\,.
\label{eq:l3l}
\eeq
For $\mu \rightarrow e$ conversion a similar relation holds:
\beq
CR(\mu\textrm{N}\rightarrow e\textrm{N})\equiv \frac{\Gamma(\mu\textrm{N}\rightarrow e\textrm{N})}{\Gamma_{capt}}
 = \frac{16\alpha^4 Z}{\Gamma_{capt}}Z_{eff}^4 \left| F(q^2)\right|^2 BR(\mu \rightarrow e\gamma )\,,
\label{eq:meN}
\eeq
where $Z$ is the proton number of the nucleus $N$, $Z_{eff}$ is an effective atomic charge obtained
by averaging the muon wave function over the nuclear density~\cite{zeff}, $F(q^2)$ denotes the nuclear
form factor at momentum transfer $q$~\cite{nucff} and 
$\Gamma_{capt}$ is the measured total muon capture rate~\cite{Suzuki:1987jf}. 
In this work we consider two target materials that will be used by future experiments. 
For $^{48}_{22}\textrm{Ti}$, which will be used by the proposed PRIME experiment at J-PARC~\cite{prime}, 
 $Z_{eff}=17.6$, $F(q^2 \simeq -m_{\mu}^2) \simeq 0.54$ and 
$\Gamma_{capt} = 2.590 \times 10^6 \textrm{sec}^{-1}$.
For $^{27}_{13}\textrm{Al}$, the target material for the proposed Mu2e experiment at Fermilab~\cite{mu2e}, 
$Z_{eff}=11.48$, $F(q^2 \simeq -m_{\mu}^2) \simeq 0.64$ and 
$\Gamma_{capt} = 7.054 \times 10^5 \textrm{sec}^{-1}$.

\subsection{$SO(10)$ GUTs}

As discussed in the previous section, LFV rates crucially depend on the neutrino Yukawa coupling matrix $\bfn$. 
However this matrix cannot be reconstructed from experimental data by inverting the seesaw
formula~(\ref{eq:seesaw}): $\bfn$ and ${\bf M}_N$ together depend on 18 parameters, while $\mathcal{M}_{\nu}$
contains only 9 observables. 
A common solution is to turn to GUTs where $\bfn$
is related to the known Yukawa matrices of SM fermions.

$SO(10)$ GUTs unify all SM fermions and the right-handed neutrino of each generation in a single 
{\bf 16}-dimensional spinor representation. 
The Higgs representation
assignments are determined by the following decompositions of the direct products:
\bea
 {\bf 16} \otimes {\bf 16} &=& {\bf 10} \oplus {\bf 120} \oplus {\bf 126} ,\\
 {\bf 16} \otimes {\bf \bar{16}} &=& {\bf 1} \oplus {\bf 45} \oplus {\bf 210}\,.
\eea

Many $SO(10)$ models exist in the literature with different
choices of Higgs representations and, frequently, with additional flavor symmetries. These models can
be divided in two general classes~\cite{so10rev}. The first uses only low dimensional Higgs multiplets {\bf
10, 16, 45} with some nonrenormalizable operators constructed from them. This necessarily leads to
large $R-$parity violation so that these models cannot provide a DM candidate. 
Models in the other class involve {\bf 10, 120} or {\bf 126} Higgs representations, have renormalizable couplings, 
preserve $R-$parity, and are often referred to
as minimal Higgs models.
The resulting set of sum-rules for the mass matrices are 
\bea
  \bfu v_u &=& {\bf f}_{10} v^{10}_u+ {\bf f}_{126} v^{126}_u+ {\bf f}_{120} v^{120}_u\,, \nonumber\\
  \bfd v_d &=& {\bf f}_{10} v^{10}_d+ {\bf f}_{126} v^{126}_d+ {\bf f}_{120} v^{120}_d\,, \nonumber\\
  \bfe v_d &=& {\bf f}_{10} v^{10}_d-3{\bf f}_{126} v^{126}_d+ {\bf f}_{120} v^{120}_d\,, \label{so10_relat}\\
  M_{\nu,LR} \equiv \bfn v_u &=& {\bf f}_{10} v^{10}_u-3{\bf f}_{126} v^{126}_u+ {\bf f}_{120} v^{120}_u\,, \nonumber\\
  M_{\nu,RR} \equiv {\bf M}_N &=& {\bf f}_{126} V_R \,,\nonumber\\
  M_{\nu,LL} &=& {\bf f}_{126} v_L \nonumber\,,
\eea
where ${\bf f}_\mathcal{R}$ are $SO(10)$ Yukawa coupling matrices, $v^\mathcal{R}_{u,d}$ are VEVs of the various
$SU(2)_L$ doublets
(with Higgs fields residing in $\mathcal{R} \equiv {\bf 10},\ {\bf 126},\ {\bf 120}$), and 
$v_L$ and $V_R$ are the $B-L$ breaking VEVs of the $SU(2)_L$ triplet and singlet respectively. In the
type-I seesaw, which we consider in this paper, $v_L =0$ and SUSY prevents it reappearance via loop
diagrams~\cite{seesaw_stab}.

From Eq.~(\ref{so10_relat}), if Higgs superfields reside in ${\bf 10}$, as they do in the simplest scenarios,
then the neutrino Yukawa matrix will be identical to the up-quark Yukawa at $M_{GUT}$. 
If the higgses are in ${\bf 126}$, then $\bfn = -3 \bfu$. 
A dominant ${\bf 120}$ Higgs would lead to at least a pair of degenerate heavy up-quarks~\cite{Mohapatra:1979nn}
and thus is phenomenologically excluded.
Motivated by the above, we introduce a {\it {neutrino Yukawa unification parameter}} as 
\begin{equation}
\bfn \propto R_{\nu u} \bfu\,,
\label{yukpar}
\end{equation}
and consider two cases,\footnote{Henceforth, we denote $|R_{\nu u}|$ by $R_{\nu u}$.} 
$|R_{\nu u}|=1$ and $|R_{\nu u}|=3$. 
Note that $\bfn$ and $\bfu$ need not be aligned: the subdominant contributions from 
other higgs multiplets and/or flavor group structure can lead to different diagonalization matrices. 
To keep our discussion as simple as possible, we consider two extreme cases of the mixing present in $\bfn$.

{\bf Large mixing:} The measured values of the neutrino mixing angles (\ref{nuoscil}) are consistent
with the so-called tri-bimaximal pattern~\cite{tribimix,Harrison:2002kp}, where  
$\sin^2 \theta_{12}=\frac{1}{3},\ \sin^2 \theta_{23}=\frac{1}{2},\ \sin^2\theta_{13}=0$.
Thus it is reasonable to postulate that mixing in the neutrino Yukawa matrix at the GUT scale also has a tri-bimaximal
form. In other words, we assume that the observed MNS mixing matrix arises only from 
the left-handed rotation matrix ({\it i.e.}, we set ${\bf V_{\nu_R}}=\mathds{1}$).
We take a neutrino Yukawa matrix 
of the form\footnote{This corresponds to trivial misalignment matrix ${\bf R}=\mathds{1}$ in the Casas-Ibarra
parametrization~\cite{casas}.}
\beq
 \bfn =R_{\nu u}{\bf V_{\nu_L}}\bfu^{\rm {diag}}\,,
\label{eq:mnscase}
\eeq
with
\beq
\label{eq:hs}
{\bf V^\dagger_{\nu_L}}={\bf U^\dagger_{\nu}}=
 \begin{pmatrix}
 \sqrt{\frac{2}{3}}c_{\chi}c_{\phi}+i\sqrt{\frac{2}{3}}s_{\chi}s_{\phi} & \frac{1}{\sqrt{3}} &
 -\sqrt{\frac{2}{3}}c_{\chi}s_{\phi}-i\sqrt{\frac{2}{3}}s_{\chi}c_{\phi} \\
 -\frac{c_{\chi}c_{\phi}+is_{\chi}s_{\phi}}{\sqrt{6}}
 -\frac{c_{\chi}s_{\phi}-is_{\chi}c_{\phi}}{\sqrt{2}} & \frac{1}{\sqrt{3}} &
 -\frac{c_{\chi}c_{\phi}-is_{\chi}s_{\phi}}{\sqrt{2}}
 +\frac{c_{\chi}s_{\phi}+is_{\chi}c_{\phi}}{\sqrt{6}} \\
 -\frac{c_{\chi}c_{\phi}+is_{\chi}s_{\phi}}{\sqrt{6}}
 +\frac{c_{\chi}s_{\phi}-is_{\chi}c_{\phi}}{\sqrt{2}} & \frac{1}{\sqrt{3}} &
  \frac{c_{\chi}c_{\phi}-is_{\chi}s_{\phi}}{\sqrt{2}}
 +\frac{c_{\chi}s_{\phi}+is_{\chi}c_{\phi}}{\sqrt{6}}
 \end{pmatrix}\,,
\eeq
where $s_{\chi}=\sin\chi,\ c_{\chi}=\cos\chi,\ s_{\phi}=\sin\phi,\ c_{\phi}=\cos\phi$ 
and the parameters $\chi,\ \phi\in [0,2\pi]$. 
This is the simplest generalization of a tri-bimaximal mixing ($\chi=\phi=0$) that allowes CP violation~\cite{Harrison:2002kp}. 
A fit to experimental values (\ref{nuoscil}) reveals that the Harrison-Scott parameters $\chi$ and $\phi$ are restricted to the vicinity of 
$\chi+\phi \simeq n\pi$, with the upper bound on $\theta_{13}$ imposing the strongest constraint. 
Ignoring RGE effects, we invert the seesaw formula and obtain the approximate form for
 the Majorana mass matrix
\beq
{\bf M}_N \simeq {\rm diag}\left(\frac{m^2_u}{m_{\nu_1}},\ \frac{m^2_c}{m_{\nu_2}},\
\frac{m^2_t}{m_{\nu_3}}\right) \times R_{\nu u}^2\,.
\label{eq:mnsRHN}
\eeq
We will consider this case as representative of the large mixing scenario.

{\bf Small mixing:} For the small mixing scenario, we take the neutrino and up-quark Yukawa matrices to be exactly aligned with
each other at the GUT scale,
\beq
 \bfn =R_{\nu u}\bfu\,,
\label{eq:ckm}
\eeq
so that neutrino mixing is given by the CKM matrix. Then, in the absence of significant RGE magnification 
effects, the Majorana mass matrix cannot be diagonal.  
Assuming tri-bimaximal mixing in $\mathcal{M}_{\nu}$ and neglecting the small mixing in $\bfn$ we can
estimate eigenvalues ${\bf M}_N$ for the normal hierarchy of light neutrinos 
($m_{\nu_1}\ll m_{\nu_2} \ll m_{\nu_3}$),
\beq
 M_{N_1}\simeq\frac{3m_u^2}{m_{\nu_2}}R_{\nu u}^2\,,\ 
 M_{N_2}\simeq\frac{2m_c^2}{m_{\nu_3}}R_{\nu u}^2\,,\ 
 M_{N_3}\simeq\frac{m_t^2}{6m_{\nu_1}}R_{\nu u}^2\,. 
\label{eq:ckmRHN}
\eeq
For the inverted mass hierarchy ($m_{\nu_1}\simeq m_{\nu_2} \gg m_{\nu_3}$), a similar procedure yields
\beq
 M_{N_1}\simeq\frac{3m_u^2}{m_{\nu_2}}R_{\nu u}^2\,,\ 
 M_{N_2}\simeq\frac{2m_c^2}{3m_{\nu_1}}R_{\nu u}^2\,,\ 
 M_{N_3}\simeq\frac{m_t^2}{2m_{\nu_3}}R_{\nu u}^2\,. 
\label{eq:ckmRHN2}
\eeq
Notice that the largest RHN mass is controlled by the smallest light neutrino mass. 

From Eqs.~(\ref{eq:mnsRHN})-(\ref{eq:ckmRHN2}), we see that RHNs have a very strong mass hierarchy ``quardatic''
to the one in up-quark sector: $M_{N_1}:M_{N_2}:M_{N_3}\sim m_u^2 : m_c^2 : m_t^2$. For this reason, only the
spectrum with normal hierarchy of light neutrinos is feasible. 
A quasi-degenerate spectrum ($m_{\nu_1}\simeq m_{\nu_2}\simeq m_{\nu_3}$) would require the lightest
Majorana mass to be in the $10^2-10^3$~GeV range with significant L-R mixing in the sneutrino sector
which is in confict with our approximations (see the discussion pertaining to Eq.~\ref{eq:snumass}). 
Moreover, such light Majorana masses make successful thermal leptogenesis impossible. 
The inverse hierarchical case  
would require the heaviest Majorana mass to be of order $10^{17}$~GeV, which is well above the GUT scale. 
This type of spectrum also suffers instabilities under very small changes to ${\bf M}_N$ and RGE
evolution~\cite{Albright:2004kb}. 
In the rest of the paper we concentrate on the normal hierarchy case, which is also 
favored by GUT model building~\cite{Albright:2009cn}.

\section{Procedure and results}
\label{sec:results}

We extensively modified ISAJET~\cite{isajet} by including the neutrino sector and by implementing RGE evolution in matrix form 
to incorporate flavor effects in both the quark and lepton sectors. 
The resulting program ISAJET-M performs RGE evolution in the MSSM augmented with RHNs at 2-loop level taking into account various thresholds
and computes sparticle spectra including radiative corrections. 
The computation of the neutralino relic density is done using the IsaReD code~\cite{isared} and LFV rates are
computed using the full one-loop formulae from Ref.~\cite{hisano}. 
A graphical outline of our procedure is presented in Fig.~\ref{fig:evolchart},
and details of the program are provided in Appendix~\ref{app:code}.
\FIGURE[!t]{
\epsfig{file=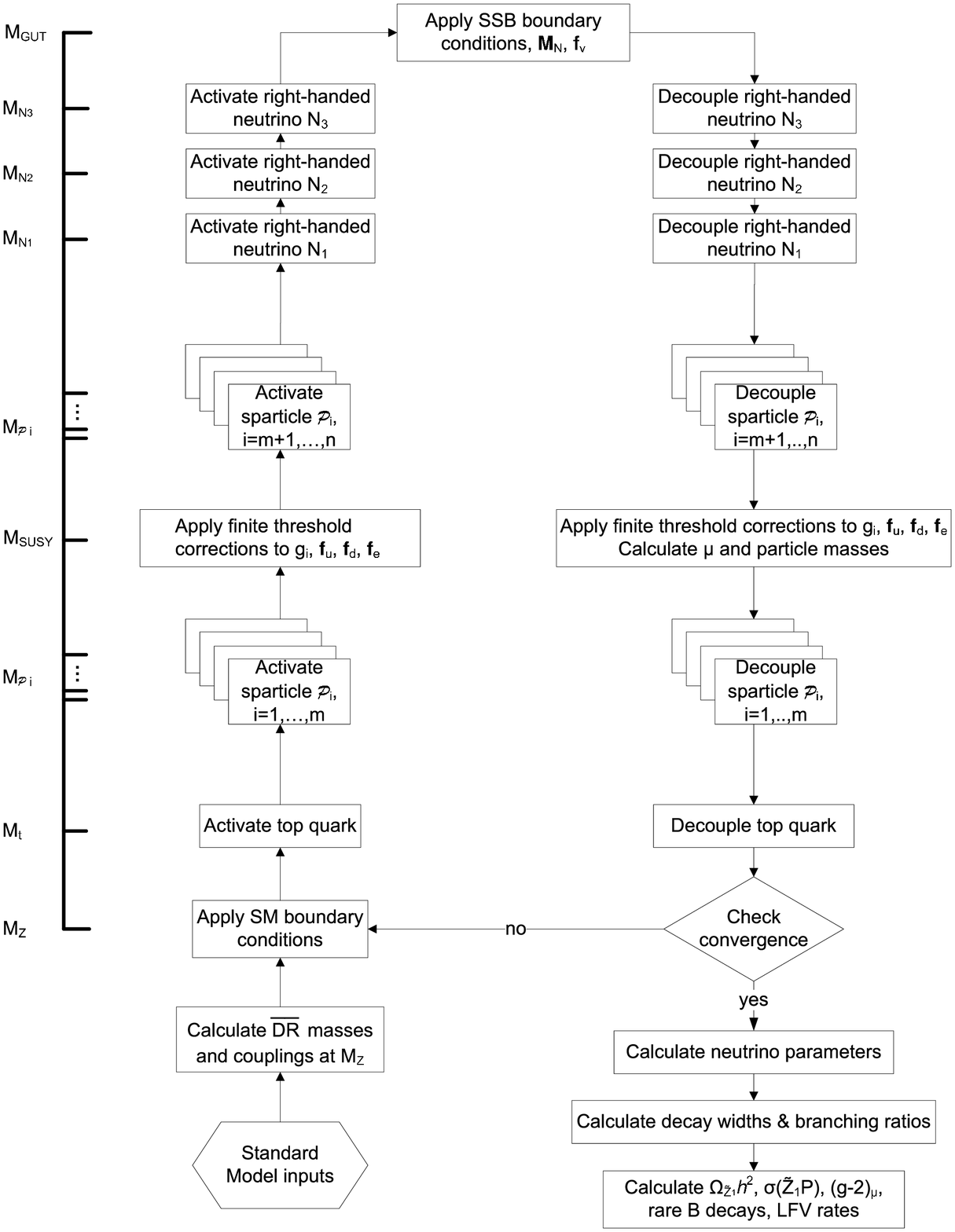,width=14.5cm,angle=0} 
\caption{
Code flowchart.
$\mathcal{P}_i$ represent sparticles and Higgses arranged in the ascending order of their masses. 
The code is described in Appendix~\ref{app:code}.}
\label{fig:evolchart}
}

In the neutrino sector we employ a ``top-down'' approach in which $\bfn$ and ${\bf M}_N$ are
inputs at the scale $M_{GUT}$. Physical neutrino masses and mixings are derived results which we
require to be within the experimental bounds~(\ref{nuoscil}). 
We consider the two cases for neutrino Yukawa unification parameter, $R_{\nu u}=1$ and $R_{\nu u}=3$ that were introduced earlier. 
For each case we consider scenarios with
large and small neutrino mixings using Eqs.~(\ref{eq:mnscase}) and (\ref{eq:ckm}) to set the neutrino Yukawa matrix at
$M_{GUT}$. 
This also restricts Majorana masses to be below that scale, {\it i.e.} ${\rm max}(M_{N_i})\lesssim M_{GUT}$.

In the quark sector we choose a basis at the weak scale in which CKM rotation is entirely in the up-quark
sector: we set the fermion rotation matrices (\ref{eq:fermass}) to be ${\bf V_{u_L}}={\bf V_{u_R}}={\bf
V}_{CKM}$ and ${\bf V_{d_L}}={\bf V_{d_R}}=\mathds{1}$.
Note that this does not mean that $\bfd$ remains diagonal at all scales -- off-diagonal terms will be
generated at higher scales due to RGE effects. 
Similarly, for charged leptons we set ${\bf V_{e_L}}={\bf V_{e_R}}=\mathds{1}$ at $M_Z$.

Regarding the SUSY sector we work in the well-studied scenario specified by the parameter set,
\beq
 m_0,\ m_{1/2},\ A_0,\ \tan\beta ,\ sgn(\mu)\,,
\label{msugra}
\eeq
where GUT-scale boundary conditions are universal and defined by Eq.~(\ref{eq:univbc}).
This choice of boundary conditions is frequently referred to as mSUGRA-like. 
Instead of scanning over the full parameter space which would be exceedingly computational intensive, 
we study specific points for each DM-allowed region. 
Throughout this work we take $\mu >0$ as suggested by measurements of the muon anomalous
magnetic moment~\cite{g-2,susyg-2,isaamu} and set the pole mass of top quark $m_t=171$~GeV in accord with Tevatron
data~\cite{topmass}. 
For the DM relic density, we consider the conceptually simplest scenario in which the DM is comprised only of
the lightest neutralino $\tz_1$ that is thermally produced in the standard $\Lambda$CDM cosmology. 
We first calculate the neutralino Relic Density (RD) and LFV rates in our framework {\it with} SUSY-seesaw using 
points from Table~\ref{tab:msugra} that have WMAP-allowed values in the mSUGRA framework {\it without} seesaw --
a procedure commonly used in the literature. 
Then, if the RD turns out to be too high, as is frequently the case, 
we find new points consistent with the WMAP
range (\ref{eq:wmap}) by adjusting SSB parameters and calculate the corresponding LFV rates.

\subsection{Large mixing}
\label{sec:lmix}

We begin by considering the $R_{\nu u}=1$ case.
Numerically we find that the GUT-scale Majorana mass matrix\,,
\beq
{\bf M}_N = {\rm diag}\left( 4.75\times 10^{-6},4.75\times 10^{-5},1\right) \cdot 1.398\times 10^{14}\,\text{GeV}\,,
\label{mmaj}
\eeq
produces 
the spectrum $m_{\nu_1}\sim 10^{-5}$~eV, $m_{\nu_2}\simeq\sqrt{\Delta m^2_{21}}\simeq 8\times 10^{-3}$~eV
and $m_{\nu_3}\simeq\sqrt{\Delta m^2_{31}}\simeq 5\times 10^{-2}$~eV 
that is in accord with experimental limits (\ref{nuoscil}). 
We chose the mass of the lightest RHN to be far above 
$M_{SUSY}$ to prevent the unwanted mixing in the sneutrino sector.
Equation~(\ref{mmaj}) is in good agreement with our estimate (\ref{eq:mnsRHN}) up to a factor of
$\sim 2$ reduction of up-quark Yukawa couplings (see for example Ref.~\cite{Barger:1992ac}) due to the RGE effects. 
This is because $\bfn$, ${\bf M}_N$ and the spectrum of light neutrinos are hierarchical and as such 
experience very little change in RGE evolution~\cite{Antusch:2005gp}.

\FIGURE[!t]{
\epsfig{file=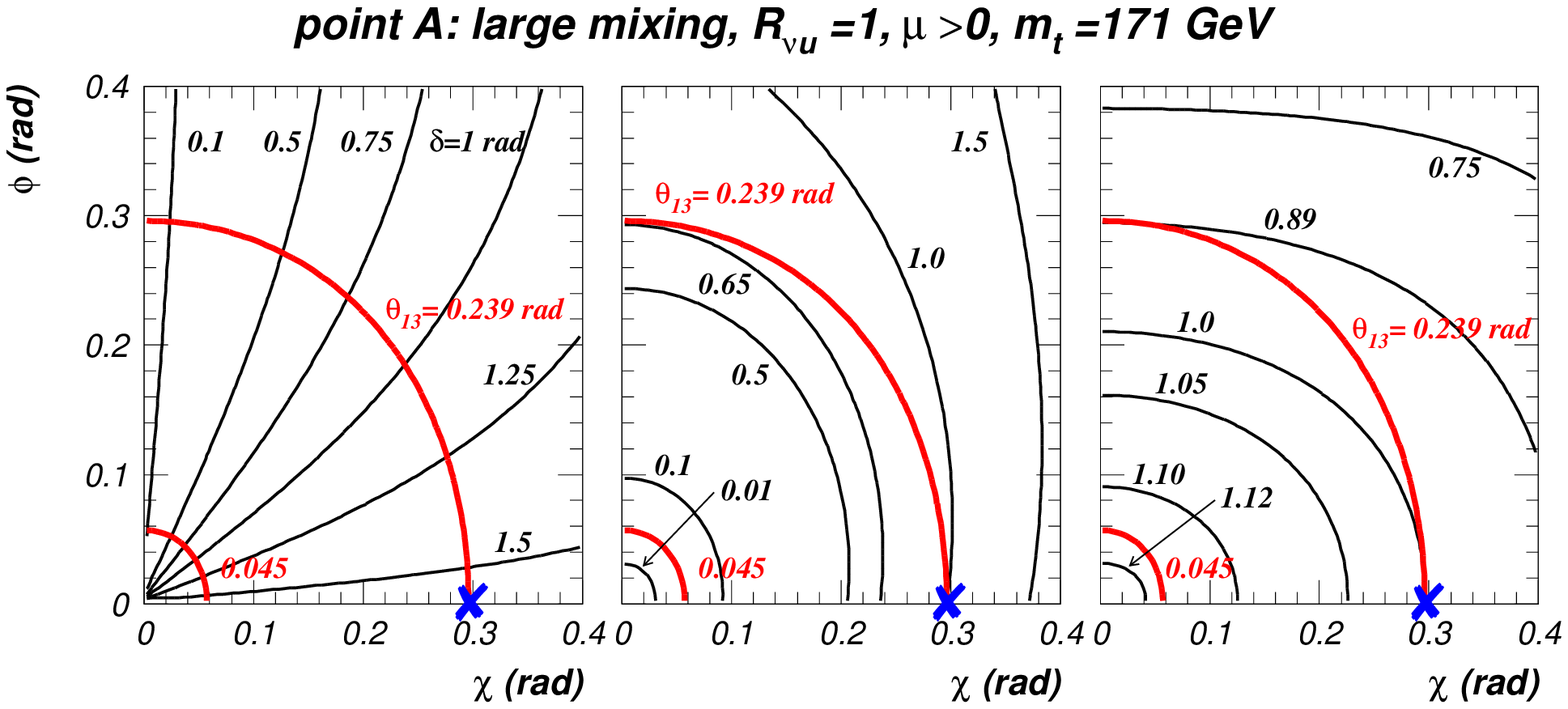,width=15.5cm,angle=0} 
\caption{\label{fig:hs_a} 
Dependence of the weak-scale Dirac phase (left), $BR(\mu\rightarrow e\gamma)$ (middle) and  
$BR(\tau\rightarrow \mu\gamma)$ (right) on GUT-scale values of the Harrison-Scott parameters for 
benchmark point A with Yukawa unification parameter $R_{\nu u}=1$. 
The contours for LFV branching ratios represent enhancement factors with respect to the point with $\delta=\pi/2$,
$\{\phi,\,\chi\}=\{0,\,0.294\}$ that is marked by the blue cross; all angles are in radians. 
The thick red lines are iso-$\theta_{13}$ contours for  $\theta_{13}=0.239$ (at the CHOOZ bound~\cite{chooz}; see Eq.~\ref{nuoscil}) and 
for the ultimate 90\% C.~L. reach of the Daya Bay experiment $\theta_{13}= 0.045$~\cite{dayabay}.
The contours remain essentially unchanged for all other points with large mixing and for $R_{\nu u}=3$.
}}

In the Harrison-Scott parameterization (\ref{eq:hs}), that we use for the mixing in the neutrino Yukawa matrix at the GUT scale,
the Dirac phase $\delta$ and mixing angles are a function of $\phi$ and $\chi$. $\delta=\pi/2$ 
for $\phi=0$ -- see the left frame of Fig.~\ref{fig:hs_a}. 
It is known that LFV rates depend on the value of the unknown Dirac phase~\cite{arganda}.
In addition, most LFV rates are very sensitive to the value of $\theta_{13}$~\cite{lfv_th13} for which
only an upper bound exists.
To quantify these dependences, in the middle frame of Fig.~\ref{fig:hs_a} we show the branching ratio for 
$\mu\rightarrow e\gamma$ as a function of the Harrison-Scott parameters. 
We present them as enhancement factors relative to the rates at
$\{\phi,\,\chi\}=\{0,\,0.294\}$ for which $\theta_{13}=0.239$ (or $\sin^2 \theta_{13}=0.056$) 
and $\delta=\pi/2$. We see that with  $\theta_{13}$ fixed, the branching ratio changes by up to $\sim 35\%$ 
under variation of $\delta$. The dependence on $\theta_{13}$ is greater and more complex: 
rates change by about two orders of magnitude for $\theta_{13}$ ranging from 0.239 to 0.045 
(or $\sin^2 2\theta_{13}=0.008$), with the latter being the ultimate reach of the Daya Bay experiment~\cite{dayabay}.
Closer to $\phi=\chi=0$, the rates change much faster and drop by several more orders of magnitude.
The rates for the other $\theta_{13}$-dependent LFV processes follow the same pattern as expected from 
Eqs.~(\ref{eq:l3l}) and (\ref{eq:meN}). 
On the other hand, rates for $\tau\rightarrow \mu \gamma$ and $\tau\rightarrow 3\mu$ are relatively independent of 
$\theta_{13}$~\cite{lfv_th13}. 
In the right frame of Fig.~\ref{fig:hs_a} we show enhancement factor contours for $\tau\rightarrow \mu \gamma$.
We see that rates change with $\theta_{13}$ by only $\sim 13\%$: 
this variation is an artifact of the paramterization (\ref{eq:hs}), in which $\sin^2
\theta_{12}=0.33/\cos^2\theta_{13}$. Also variation with respect to $\delta$ is smaller -- only up to $\sim 10\%$. 
We numerically confirmed that $\tau\rightarrow 3\mu$ rates behave similarly, as expected from Eq.~(\ref{eq:l3l}).

In Figs.~\ref{fig:meg_mns}, \ref{fig:taulg_mns}, \ref{fig:l3l_mns} and \ref{fig:mne_mns}, we show the LFV rates 
along with current experimental bounds (horizontal dashed lines) and projected future sensitivities
(dash-dotted lines). 
To account for the aforementioned dependences on $\theta_{13}$ and $\delta$, we show rates for 
$\{\phi,\,\chi\}=\{0,\,0.294\}$ (for which $\sin^2\theta_{13}=0.056$ and $\delta=\pi/2$) as diagonally hatched bars and for 
$\phi=\chi=0$ (resulting in $\sin^2\theta_{13}=0$) as solid bars. 
We also present rates for $\{\phi,\,\chi\}=\{0,\,0.022\}$ (giving $\sin^2\theta_{13}=0.002$ and $\delta=\pi/2$) as cross-hatched bars to
indicate the upper limit on the rates if the Daya Bay~\cite{dayabay} and Double Chooz~\cite{dchooz} experiments produce a null result. 

In the bulk~\cite{bulk,Baer:1995nc} and the stau-coannihilation~\cite{dm:stau,isared} 
regions, the neutralino RD is within the WMAP range due to $\tz_1$ 
interactions with the the lighter stau $\ttau_1$. Under universal boundary conditions $\ttau_1$ is a
dominantly right-handed state and as such remains unaffected by the neutrino Yukawa coupling\footnote{A detailed discussion of
neutrino Yukawa coupling effects on SUSY spectrum and DM observables can be found in Ref.~\cite{msugrarhn}.}. 
Therefore, the use of mSUGRA values (points A and B) in models with RHNs still produces the correct RD. At point A in the 
$R_{\nu u}=1$ case,
$BR(\mu\rightarrow e\gamma)$ changes from $1.77\times 10^{-16}$ for $\phi=\chi=0$, to $8.57\times 10^{-8}$ for
$\{\phi,\,\chi\}=\{0,\,0.294\}$; intermediate allowed $\phi$ and $\chi$ values produce rates between these values.
For $\tau\rightarrow \mu \gamma$ and $\tau\rightarrow 3\mu$, the dependence on $\phi$ and $\chi$ parameters is reversed: 
larger $\phi$ and $\chi$ produce smaller rates. For example, at point A in the $R_{\nu u}=1$ case, the
$\tau \rightarrow \mu\gamma$ branching fraction is $2.09\times 10^{-7}$ for $\phi=\chi=0$ and is $1.77\times
10^{-7}$ for $\{\phi,\,\chi\}=\{0,\,0.294\}$. 
We see that $\tau\rightarrow \mu \gamma$ and $\tau\rightarrow 3\mu$ are excellent probes of LFV: the current experimental bound of 
$\tau\rightarrow \mu \gamma$ rules out the bulk region for all values of $\phi$ and $\chi$.

\FIGURE[t]{
\epsfig{file=meg_mns.eps,width=15cm,angle=0} 
\caption{\label{fig:meg_mns} 
Radiative LFV decay rates in the large mixing case for two values of $R_{\nu u}$ 
(c.f. Eqs.~\ref{eq:llog} and \ref{eq:ms}). 
The heights of the solid bars show rates for exact tri-bimaximal mixing $\chi=\phi=0$ at the GUT scale.
Diagonally hatched bars represent  $\{\phi,\,\chi\}=\{0,\,0.294\}$ yielding the maximum allowed $\theta_{13}$ and
$\delta=\pi/2$.  
Cross-hatched bars represent $\{\phi,\,\chi\}=\{0,\,0.022\}$ which has $\delta=\pi/2$ and
$\theta_{13}\simeq 0.045$ that is the ultimate reach of the Daya Bay experiment. 
Dashed lines represent the current bound and dash-dotted lines the projected sensitivity 
listed in Table~\ref{tab:lfv}. The letters denote various benchmark points presented in 
Tables~\ref{tab:msugra}, \ref{tab:mns} and \ref{tab:mns3fu}. Points C through G have a RD above the WMAP bound. 
The corresponding WMAP-consistent points indicated by
subscripts $s$, $t$ and $g$ are obtained by adjusting $m_0$, $A_0$ and $m_{1/2}$, respectively.
}}
\FIGURE[tbh]{
\epsfig{file=tmg_mns.eps,width=15cm,angle=0} 
\epsfig{file=teg_mns.eps,width=15cm,angle=0} 
\caption{\label{fig:taulg_mns} 
Similar to Fig.~\ref{fig:meg_mns} for $\tau$ radiative LFV decays. 
For $\tau\rightarrow\mu\gamma$, the $\{\phi,\,\chi\}=\{0,\,0.022\}$ case  is not shown as it
produces rates very close to those for $\{\phi,\,\chi\}=\{0,\,0.294\}$ (cross-hatched).
}}
\FIGURE[tbh]{
\epsfig{file=m3e_mns.eps,width=13.5cm,angle=0} 
\epsfig{file=t3m_mns.eps,width=13.5cm,angle=0} 
\epsfig{file=t3e_mns.eps,width=13.5cm,angle=0} 
\caption{\label{fig:l3l_mns}
Similar to Fig.~\ref{fig:meg_mns} for trilepton LFV decays.
}}
\FIGURE[tbh]{
\epsfig{file=meTi_mns.eps,width=15cm,angle=0} 
\epsfig{file=meAl_mns.eps,width=15cm,angle=0} 
\caption{\label{fig:mne_mns}
Similar to Fig.~\ref{fig:meg_mns} for LFV $\mu\rightarrow e$ conversion rates 
in titanium (upper frames) and aluminum (lower frames) targets. 
}}
%

In the stop-coannihilation region~\cite{dm:stop} the picture is radically different.  
A naive use of input parameters for
point C gives a neutralino relic density $\Omega_{\tz_1}h^2=0.34$, well above the WMAP bound. This is
because $\ttop_1$ is pushed to a higher mass causing stop-coannihilation to cease. 
To restore the stop-coannihilation mechanism and bring $\Omega_{\tz_1}h^2$ down to the WMAP range, 
one can counteract the effect of $\bfn$ by adjusting SSB parameters. 
Adjusting the common scalar mass parameter $m_0$, with the rest of SSB parameters held fixed, can lower the stop mass back
to the desired mass range
leading to a new WMAP-consistent point that we denote\footnote{Hereafter we
use subscripts $s,\ t$ and $g$ to denote points obtained from those in Table~\ref{tab:msugra}  
by adjusting the value of one the model parameters: {\it s}calar mass $m_0$, {\it t}rilinear coupling $A_0$ or
{\it g}augino mass $m_{1/2}$, respectively.
Single and double primes are used to further distinguish modified points in the large mixing with $R_{\nu u}=3$ and the
small mixing with $R_{\nu u}=1$ scenarios.} 
$C_s$ with parameter values shown in Table~\ref{tab:mns}.
Increasing $m_0$ also makes sleptons lighter and increases their mixing, as can be seen from
Eqs.~(\ref{eq:llog}) and (\ref{eq:ms}), leading to a factor of $\sim 2.8$ increase in LFV rates as compared to point C.
The stop mass can also be lowered by dialing the trilinear A-term,
resulting in another modified point $C_t$. 
The rates increase with respect to values at point C by only about 10\%. 
Alternatively, one can raise the $\tz_1$ mass closer to that of the $\ttop_1$ by adjusting the common gaugino mass parameter
$m_{1/2}$ producing a correct RD at the point $C_g$.
Since the required increase is small, rates again grow only by $\sim 10\%$.

\TABLE{
\begin{tabular}{c||cccc|c}
\hline
Point & $m_0$ & $m_{1/2}$ & $A_0$ & $\tan\beta$ & Region \\
\hline
$C_s$ & {\bf 94} & 300 & -1095 & 5  & $\ttop$-coan.\\
$C_t$ & 150 & 300 & {\bf -1120} & 5  & $\ttop$-coan.\\
$C_g$ & 150 & {\bf 294} & -1095 & 5  & $\ttop$-coan.\\
$D_s$ & {\bf 440} & 450 & 0    & 51 & $A$-funnel \\
$D_t$ & 500 & 450 & {\bf 150}  & 51 & $A$-funnel \\
$D_g$ & 500 & {\bf 724} & 0    & 51 & $A$-funnel \\
$E_s$ & {\bf 1722} & 300 & 0   & 10 & HB/FP \\
$F_s$ & {\bf 3607} & 1000 & 0  & 10 & HB/FP \\
$F_g$ & 3143 & {\bf 806.5} & 0  & 10 & HB/FP \\
$G_s$ & {\bf 2435} & 130 & -2000 & 10 & $h$-funnel \\
$G_t$ & 2000 & 130 & {\bf -1680} & 10 & $h$-funnel \\
\hline
\end{tabular}
\caption{Modified benchmark points for mSUGRA-seesaw in case of the large mixing with $R_{\nu u}=1$
 obtained from their counterparts in Table~\ref{tab:msugra} by adjusting the parameter highlighted in boldface
to produce the RD dictated by WMAP. 
All dimensionful parameters are in GeV. For all points $\mu >0$ and $m_t = 171$~GeV.}
\label{tab:mns}}

In the A-funnel region~\cite{Baer:1995nc,Afunnel}, $\bfn$ pushes the mass of the CP-odd Higgs boson $A$ up and away from
the resonance, resulting in a larger RD of 0.144 for point D. 
This can be reduced by decreasing the scalar mass parameter (point $D_s$) 
or increasing the value of the trilinear A-term (point $D_t$), 
which lower the $A$ mass back to the resonance regime. In both cases LFV rates increase by about 8\%.
Increasing the gaugino mass parameter can increase the $\tz_1$ mass to the resonance value $m_{\tz_1}=0.5m_A$. 
But the $A$ mass also grows with $m_{1/2}$ although slower than the neutralino mass, 
so a large dialing is required, resulting in point $D_g$.
This large change in $m_{1/2}$ increases the masses of the charginos in the loop resulting in about 
an order of magnitude drop in LFV rates.

In the lower part of the HB/FP region~\cite{dm:fp}, at point E the RD is two orders of magnitude above the WMAP
range. This is due to an increased value of $\mu$ from the $\bfn$ effect that can be counteracted by increasing
$m_0$, resulting in new WMAP-allowed point $E_s$.
The heavier sfermion spectrum causes LFV rates to decrease by about a factor of two. 
Adjustment of the trilinear parameter can somewhat lower $\mu$~\cite{susy_books}, but not enough to get back into the HB/FP regime. 
It would be possible to reduce the RD by lowering $m_{1/2}$ to 189~GeV, but at this value the chargino mass falls
below the LEP2 bound of 103.5~GeV~\cite{lep2}.

In the upper portion of the HB/FP region, the neutrino Yukawa couplings have an extremely large effect -- the RD
at point F becomes 12.3. Similarly to point E, the RD can be lowered by increasing the scalar mass parameter.
In this part of the HB/FP region the chargino mass is sufficiently high that $m_{1/2}$ can be lowered without violating the LEP2 chargino
bound, resulting in a WMAP-allowed value at point $F_g$.
Since charginos become lighter with this adjustment, LFV rates increase by about 40\%.

In the light Higgs resonance region~\cite{Baer:1995nc,hfunnel}, the neutrino Yukawa coupling also destabilizes the $\tz_1$
annihilation mechanism producing too large a RD at point G. Although neither the $\tz_1$ nor $h$ masses are moved away
from the desired regime $2m_{\tz_1}=m_h$, the resonance mechanism ceases to function because $\tz_1$ becomes bino-like and can no
longer couple to the higgs. The neutralino-higgs coupling can be restored by lowering $\mu$ by increasing
the scalar mass. This decreases LFV rates by about 20\%.
The desired value of the neutralino-higgs coupling can also be achieved by adjustment of the trilinear parameter
resulting in point $G_t$. In this case, the LFV rates increase only marginally compared to point G.

For $R_{\nu u}=3$, we find that experimental limits (\ref{nuoscil}) are satisfied  
for a GUT-scale Majorana mass matrix of the form,
\beq
{\bf M}_N = {\rm diag}\left( 4.5\times 10^{-6},4.5\times 10^{-5},1\right) \cdot 1.3\times 10^{15}\,\text{GeV},
\label{mmaj3}
\eeq
which we use in subsequent computations. 
This is a simple rescaling of Eq.~(\ref{mmaj}) as expected from the seesaw formula.
We have numerically verified that the dependence of the neutrino mixing parameters and LFV rates follow the same
pattern shown in Fig.~\ref{fig:hs_a}. Thus, rates on Figs.~\ref{fig:meg_mns}-\ref{fig:mne_mns} are presented for the
same choice of $\phi$ and $\chi$ discussed earlier.

As for $R_{\nu u}=1$, the RD in points A and B remain unaffected by the presence of additional neutrino Yukawa
couplings. However, the very large neutrino Yukawa generates large off-diagonal terms in the slepton mass matrix, as can be
seen from Eq.~(\ref{eq:llog}), that boost LFV rates by more than an order of magnitude as compared to the $R_{\nu u}=1$ case.
Nevertheless, this is still not enough to rule out point B for the whole range of $\chi$ and $\phi$ values.

At point C, the RD is too large: the $\ttop_1$ mass is pushed further away from the $\tz_1$ due to larger
neutrino Yukawa effects. Thus restoration of the stop coannihilation mechanism requires larger
adjustments of scalar mass and trilinear coupling parameters, 
leading to new WMAP-consistent points $C'_s$ and $C'_t$ listed in Table~\ref{tab:mns3fu}. 
Unlike the $R_{\nu u}=1$ case, adjustment of $m_{1/2}$ cannot restore the stop-coannihilation:
effects of $\bfn$ make $m_{\ttau_1}<m_{\ttop_1}$ and the RD is lowered to the WMAP range at $m_{1/2}=725$~GeV by the
stau-coannihilation mechanism. Increasing $m_{1/2}$ further makes $\ttau_1$ the LSP before the
stop-coannihilation regime can be reached.

\TABLE{
\begin{tabular}{c||cccc|c}
\hline
Point & $m_0$ & $m_{1/2}$ & $A_0$ & $\tan\beta$ & Region \\
\hline
$C'_s$ & {\bf 96} & 300 & -1095 & 5  & $\ttop$-coan.\\
$C'_t$ & 150 & 300 & {\bf -1197} & 5  & $\ttop$-coan.\\
$D'_s$ & {\bf 355} & 450 & 0     & 51 & $A$-funnel \\
$E'_s$ & {\bf 6061.5} & 300 & 0  & 10 & HB/FP \\
$F'_s$ & {\bf 7434} & 1000 & 0   & 10 & HB/FP \\
$G'_s$ & {\bf 6530} & 130 & -2000 & 10 & $h$-funnel \\
\hline
\end{tabular}
\caption{Similar to Table~\ref{tab:mns} for the large mixing and $R_{\nu u}=3$ case.}
\label{tab:mns3fu}}

In the A-funnel, larger neutrino Yukawas push $m_A$ away from the resonance resulting in $\Omega_{\tz_1}h^2=0.33$
at point D. Lowering the scalar mass parameter can bring the $A$ mass back into the resonance regime 
at point $D'_s$.
The RD can also be lowered by either adjusting $A_0$ to -692~GeV or raising $m_{1/2}$ to 745~GeV. However,
either of these bring the $\ttau_1$ mass close to the $\tz_1$ mass and activate the stau-coannihilation
mechanism; further dialing of either parameter makes $\ttau_1$ the LSP.

In the HB/FP point E, $\mu$ is pushed by neutrino Yukawa coupling to very large values resulting in large
RD, $\Omega_{\tz_1}h^2=25$. To compensate, one needs to dial the scalar mass parameter to very high values  
(point $E'_s$). 
At such a large $m_0$, sleptons become very heavy causing LFV rates to drop by about two orders of magnitude. 
Consequently, rates for LFV muon decay and $\mu -e$ conversion fall below current limits for all $\chi$ and $\phi$ values. 
For this point, $\mu$ is so large that the HB/FP regime cannot be recovered by adjusting $A_0$ or $m_{1/2}$.

Similarly, for point F we get an extremely high relic density $\Omega_{\tz_1}h^2=72$ that can be compensated 
by a very large scalar mass at point $F'_s$.
The LFV rates are lowered by a factor of $\sim 15$ so that muon LFV rates are below experimental bounds for all
allowed mixing angles. 
As in the lower part of the HB/FP region, dialing $A_0$ or $m_{1/2}$ cannot bring $\Omega_{\tz_1}h^2$ back 
in accord with WMAP.

In the $h$-resonance point G, $\mu$ is pushed so high that $\tz_1$ becomes a pure bino state unable to couple to the Higgs
boson. This can be compensated only by significantly increasing the scalar mass
yielding a new WMAP-consistent point $G'_s$ with LFV rates that are smaller 
by two orders of magnitude.

\subsection{Small mixing}

For the case of small mixing, we set $\bfn(M_{GUT})$ according to Eq.~(\ref{eq:ckm}) and 
choose the neutrino spectrum to be $m_{\nu_1}= 6\times10^{-4}$~eV, $m_{\nu_2}= 8\times 10^{-3}$~eV
and $m_{\nu_3}= 5\times 10^{-2}$~eV. Our choice for $m_{\nu_1}$ is constrained by the fact that we need $M_{N_1} \gg
M_{SUSY}$ for our approximation to remain valid. 
In this scenario LFV rates do not depend on neutrino mixing angles: Eq.~(\ref{eq:ckm}) fixes the
neutrino Yukawa matrix completely and perturbations of the structure of ${\bf M}_N$ do not produce significant changes in
RGE evolution as can be seen from Eq.~(\ref{eq:llog}).

For $R_{\nu u}=1$, full RGE evolution with our code yield the following
eigenvalues of the Majorana mass matrix:
\beq
 M_{N_1}\simeq 8\times 10^4 \,\text{GeV}\,,\ 
 M_{N_2}\simeq 3.5\times 10^9 \,\text{GeV}\,,\ 
 M_{N_3}\simeq 2.5\times 10^{15} \,\text{GeV}\,. 
\eeq
This would appear to be in conflict with thermal leptogenesis, which requires the lightest Majorana mass to be
heavier than about $10^9$~GeV~\cite{leptogen}. 
Nevertherless, successful leptogenesis is possible through the decay of the next-to-lightest RHN~\cite{DiBari}.
Also notice that  $M_{N_3}$ is closer to $M_{GUT}$ than it was
in the case of large mixing. This, combined with small mixing in $\bfn$, lead to rates that are several orders
of magnitude smaller, putting them all significantly below current experimental bounds, as shown in
Figs.~\ref{fig:llg_ckm}, \ref{fig:l3l_ckm} and \ref{fig:mne_ckm}.

\FIGURE[tbh]{
\epsfig{file=meg_ckm.eps,width=13.5cm,angle=0} 
\epsfig{file=tmg_ckm.eps,width=13.5cm,angle=0} 
\epsfig{file=teg_ckm.eps,width=13.5cm,angle=0} 
\caption{\label{fig:llg_ckm} 
Radiative LFV decay rates in the small mixing case for benchmark points presented in Tables \ref{tab:msugra} and \ref{tab:ckm}. 
All rates are below current experimental bounds.
Dash-dotted lines represent the projected future sensitivity listed in Table~\ref{tab:lfv}. 
}}
\FIGURE[tbh]{
\epsfig{file=m3e_ckm.eps,width=14cm,angle=0} 
\epsfig{file=t3m_ckm.eps,width=14cm,angle=0} 
\epsfig{file=t3e_ckm.eps,width=14cm,angle=0} 
\caption{\label{fig:l3l_ckm}
Similar to Fig.~\ref{fig:llg_ckm} for trilepton LFV decays.
}}
\FIGURE[tbh]{
\epsfig{file=meTi_ckm.eps,width=15cm,angle=0} 
\epsfig{file=meAl_ckm.eps,width=15cm,angle=0} 
\caption{\label{fig:mne_ckm}
Similar to Fig.~\ref{fig:llg_ckm} for LFV conversion rates in titanium (upper frames) and aluminum (lower frames) targets. 
}}

As discussed in Section~\ref{sec:lmix}, the use of mSUGRA values of points A and B still produces the correct RD. 
For example, at point A we have $BR(\mu \rightarrow e\gamma)=1.1\times 10^{-13}$ that is barely above
the reach of the future MEG experiment.
In the $A$-funnel, neutrino Yukawa couplings do affect the annihilation mechanism, but the effect is
small and the RD remains within the WMAP range.
For the other regions, adjustment of SSB parameters is necessary; the modified points are listed in
Table~\ref{tab:ckm}. 

\TABLE{
\begin{tabular}{c||cccc|c}
\hline
Point & $m_0$ & $m_{1/2}$ & $A_0$ & $\tan\beta$ & Region \\
\hline
$C''_s$ & {\bf 126} & 300 & -1095 & 5  & $\ttop$-coan.\\
$C''_t$ & 150 & 300 & {\bf -1106} & 5  & $\ttop$-coan.\\
$C''_g$ & 150 & {\bf 297} & -1095 & 5  & $\ttop$-coan.\\
$E''_s$ & {\bf 1505} & 300 & 0   & 10 & HB/FP \\
$F''_s$ & {\bf 3300} & 1000 & 0  & 10 & HB/FP \\
$F''_g$ & 3143 & {\bf 943} & 0  & 10 & HB/FP \\
$G''_s$ & {\bf 2205} & 130 & -2000 & 10 & $h$-funnel \\
$G''_t$ & 2000 & 130 & {\bf -1895} & 10 & $h$-funnel \\
\hline
\end{tabular}
\caption{Similar to Table~\ref{tab:mns} for the small mixing and \mbox{$R_{\nu u}=1$} case.}
\label{tab:ckm}}

Even a decoupling scale as high as $M_{N_3}\sim 10^{15}$~GeV is enough
to destabilize the stop-coannhilation mechanism, resulting in too large a RD for point C. Lowering the scalar mass parameter 
decreases the $\ttop_1$ mass to the desired RD value at point $C''_s$, leading to $\sim 50\%$ increase in the LFV rates. 
The stop coannihilation mechanism can also be restored by adjusting the trilinear A-term 
(point $C''_t$) or the gaugino mass (point $C''_g$). 
In both cases the LFV rates increase only slightly from those for point C. 
At all points $\mu \rightarrow e\gamma$ rates are slightly below the MEG reach, so this region will only
be probed by $\mu \rightarrow e$ conversion experiments.

At point E, the RD is also too high. For the reasons discussed in Section~\ref{sec:lmix}, only
adjustment of $m_0$ is possible, leading to consistency with WMAP for the values at point $E''_s$.
Since the required shift is not as significant as in the large mixing case, the LFV rates drop only by $\sim
30\%$.

In the upper part of the HB/FP region at point D, the RD exceeds the WMAP value by two orders of
magnitude. The desired higgsino content of $\tz_1$ can be restored by adjusting the scalar or gaugino mass
parameters, giving points $F''_s$ and $F''_g$.
As a result, the LFV rates change by about 10\% with respect to those at point D.

At point F, the higgsino content of $\tz_1$ is diminished by neutrino Yukawa RGE effects resulting in too high a
RD. The $h$-resonance mechanism can be restored by adjusting the scalar mass 
(point $G''_s$) or the trilinear A-term (point $G''_t$).
These adjustments change the LFV rates approximately by a factor of two with respect to the prediction for point F.

For $R_{\nu u}=3$, we find the eigenvalues of the Majorana mass matrix to be
\beq
 M_{N_1}\simeq 7\times 10^5 \,\text{GeV}\,,\ 
 M_{N_2}\simeq 3\times 10^{10} \,\text{GeV}\,,\ 
 M_{N_3}\simeq 2.2\times 10^{16} \,\text{GeV}\,. 
\eeq
The heaviest Majorana mass value is very close to $M_{GUT}\simeq 2.3\times 10^{16}$~GeV.
Because of this, the effect of $\bfn$ on RD is negligible and WMAP-consistent values are obtained for the mSUGRA points in
Table~\ref{tab:msugra}. From the resultant LFV rates in Figs.~\ref{fig:llg_ckm}-\ref{fig:mne_ckm}, 
we see that the larger neutrino Yukawa coupling produces LFV rates that are smaller by almost an order of magnitude. 
This is opposite to what we saw in the large mixing scenarios where $R_{\nu u}=3$ rates were more than an order of
magnitude greater that their $R_{\nu u}=1$ cousins.
This is also a direct consequence of the proximity of $M_{N_3}$ to $M_{GUT}$: the largest neutrino Yukawa
decouples almost immediately and off-diagonal elements in the slepton doublet matrix are generated by the much smaller
Yukawas of the first and second generations.

\section{Discussion}
\label{sec:disc}

We perfomed a detailed study of the LFV rates in the RD-allowed benchmark points, 
and demonstrated that the interconnection between the neutrino sector and neutralino dark matter is very important for
predictions of LFV rates. Proper consideration of these effects change LFV rates by
factor of a few to up to two orders of magnitude.
We emphasize that although we used $SO(10)$ models to set the structure of the neutrino Yukawa matrix
at the GUT scale, our results are generic; they hold in any type-I SUSY-seesaw scenario with 
large neutrino Yukawa couplings.

The results in Section~\ref{sec:results} imply the following about models with
universal (or mSUGRA-like) SSB boundary conditions stipulated at the GUT scale:
\begin{itemize}

\item The small mixing scenario is completely consistent with present experimental bounds on LFV. Upcoming $\mu\rightarrow
e\gamma$ experiments will probe only a very small corner of parameter space where both $m_0$ and
$m_{1/2}$ are small and $R_{\nu u}=1$. 
Future $\mu \rightarrow e$ conversion experiments, although suppressed by a factor $\sim Z\alpha /\pi$ with
respect to $\mu\rightarrow e\gamma$, have better prospects due to the very well defined experimental signal.
The PRIME experiment will be able to probe the entire bulk and stop-coannihilation regions as well as a 
significant portion of the $A$-funnel region, while the Mu2e experiment will only be able to probe the bulk and
stop-coannihilation regions for $R_{\nu u}=1$.

\item Contrary to naive expectations, in the small mixing case, order of magnitude {\it smaller} LFV rates are expected for $R_{\nu u}=3$
than for $R_{\nu u}=1$. Such small rates will only be probed by the PRIME $\mu \to e$ conversion
experiment with a $Ti$ target.

\item Future $\mu\rightarrow e\gamma$ measurements will not  rule out large mixing scenarios for any values of 
$\phi$ and $\chi$ in the mixing matrix of Eq.~(\ref{eq:hs}) because of
the high sensitivity of this channel to $\theta_{13}$. On the other hand, if $\theta_{13}$ is close to the CHOOZ bound,
only the HB/FP and $h$-funnel regions are consistent with current limits.

\item The $\tau \rightarrow \mu\gamma$ channel is an excellent probe as it is not sensitive to $\theta_{13}$. 
Current experimental limits exclude the bulk region and the
stop-coannihilation regions. For large $R_{\nu u}$,  
part of the $A$-funnel region is also excluded. Future experiments at Super Flavor factories~\cite{Bona:2007qt} 
should be able to probe the $A$-funnel region almost entirely.

\item Trilepton decays are weaker probes due to a factor $\sim \alpha$ suppression of the rates as compared to
the two body modes. Nevertheless, current data rule out $\theta_{13}$ close to the CHOOZ bound for some regions of
SUSY parameter space.

\item $\mu \rightarrow e$ conversion in nuclei is the best probe of LFV. Future experiments will have sensitivity to almost
the entire parameter space.
In the large mixing case, $\mu \rightarrow e$ conversion is highly complementary to collider searches: 
it can probe large parts of the HB/FP region which can not be probed at the LHC. 

\end{itemize}


%

The upcoming Daya Bay and Double Chooz experiments will soon be able to probe $\theta_{13}$
independently of the Dirac phase down to $\sin^2 \theta_{13}=0.002$. A signal of nonzero $\theta_{13}$ will significantly reduce
uncertainties in the predictions of LFV rates if the observed neutrino mixing arises dominantly from $\bfn$ as
in the large mixing case. The $\theta_{13}$-dependent LFV rates will be
within about two orders of magnitude of the maximum values shown in Figs.~\ref{fig:meg_mns}-\ref{fig:mne_mns} 
thus further constraining model parameter space
with current LFV data. 
For instance, the current $\mu\rightarrow e\gamma$ bound rules out a significant portion of the $A$-funnel for the case of large
mixing and either value of $R_{\nu u}$.
With future $\mu\rightarrow e\gamma$ measurements we will be able to tell if the type-I SUSY-seesaw 
can be realized in the stau-coannihilation region with
large $\bfn$ mixing regardless of $R_{\nu u}$; see point B in Fig.~\ref{fig:meg_mns}.

If the LHC finds a signal of SUSY, then GUTs become very appealing. 
One might be able to combine the knowledge of the
sparticle spectrum from the LHC with results from LFV experiments to determine the value of $R_{\nu u}$ and/or get some
information about the mixing pattern in the neutrino Yukawa. 
For example, if SUSY is found to be realized in the bulk region (point A), then a type-I SUSY-seesaw can only exist if
the mixing in $\bfn$ is small. Then measurements from the PRIME experiment could be used to identify $R_{\nu u}$.
The situation becomes even more favorable if $\theta_{13}$ is known. 
For example, if SUSY is found to be consistent with the $A$-funnel region, then with the value of $\theta_{13}$ in hand,
PRIME measurements will be able to test if the type-I SUSY-seesaw is operative for all mixing patterns and $R_{\nu u}$
values.

\section{Conclusions}
\label{sec:conclus}

In previous work~\cite{msugrarhn} we demonstrated that large neutrino Yukawa couplings can
significantly affect the neutralino relic density. This effect can be counteracted by the adjustment of SSB
parameters, with concomitant changes in the low-energy phenomenology. In this work, we studied LFV processes in
the type-I SUSY-seesaw properly taking into account neutrino Yukawa effects on the neutralino RD. For simplicity, we
considered a scenario with flavor-blind universal (or mSUGRA-like) boundary conditions defined at the GUT scale. 
In the neutrino sector we utilized the ``top-down'' approach in which neutrino Yukawa
and Majorana mass matrices are inputs at $M_{GUT}$. We considered
two cases for the neutrino-upquark unification parameter $R_{\nu u}$ (see Eq.~\ref{yukpar}) that are inspired by $SO(10)$ models. 
For each scenario we examined two extreme cases for the mixing in the neutrino Yukawa matrix. 
We found that the common practice of using WMAP allowed points of mSUGRA in models with RHNs overestimates the LFV
rates in most regions of parameter space.

In the $R_{\nu u}=1$ case we found that the neutrino-neutralino interplay can result in significant changes in
LFV predictions. 
The rates can change by a factor of up to 5 in the WMAP-allowed regions as compared to naive esimates. 
Effects are most prominent in regions with a large scalar mass parameter $m_0$ such as HP/FP and $h$-funnel regions.
If the mixing in the neutrino Yukawa matrix is small, then all LFV rates are below current experimental bounds.
In the future, this case can be probed to some extent by the MEG experiment and by the PRIME and Mu2e conversion
experiments.

The case of very large unification parameter $R_{\nu u}=3$, contrary to common lore, is not ruled out by current
bounds on LFV processes even if the mixing in the neutrino Yukawa matrix is large.
In the large mixing case a proper treatment of the neutralino-neutrino interplay leads to LFV
rates that are smaller by about two orders of magnitude than naively expected. As result, many rates fall below
current limits. Surprisingly, we found that if mixing in the neutrino Yukawa matrix is small, then for $R_{\nu u}=3$, the
LFV rates are an order of magnitude smaller than for $R_{\nu u}=1$. If this scenario is realized, then
only the future PRIME experiment will have sensitivity to some regions of the parameter space.

\section*{Acknowledgments}
We thank A.~Belyaev, S.~Blanchet, A.~Box, M.~Malinsky, H.~P\"as and X.~Tata for discussions and useful inputs, and
A.~Box and X.~Tata for forwarding us the unpublished erratum of Ref.~\cite{Castano}. VB and DM
 thank the Aspen Center for Physics for hospitality.
This work was supported by the DoE under Grant Nos. DE-FG02-95ER40896 and DE-FG02-04ER41308,
by the NSF under Grant No. PHY-0544278, and by the Wisconsin Alumni Research Foundation.

\appendix

\section{Notation and conventions}
\label{app:notation}

Here we briefly list the most relevant equations to establish our formalism; we
follow the notation and conventions of Ref.~\cite{Baer:book}.

The MSSM superpotential has the form
\beq
\hat{f}=\mu\hat{H}_u^a\hat{H}_{da}+\sum_{i,j=1,3}\left[
({\bf f}_u)_{ij}\epsilon_{ab}\hat{Q}^a_i\hat{H}_u^b\hat{U}^c_j +
({\bf f}_d)_{ij}\hat{Q}^a_i\hat{H}_{da}\hat{D}^c_j +
({\bf f}_e)_{ij}\hat{L}^a_i\hat{H}_{da}\hat{E}^c_j\right] ,
\label{eq:superpot}
\eeq
where $a,\ b$ are $SU(2)_L$ doublet indices, $i,j$ are generation indices,
$\epsilon_{ab}$ is the totally antisymmetric tensor with $\epsilon_{12}=1$ and the
superscript $c$ denotes charge conjugation.

The soft SUSY breaking part of the Lagrangian is
\bea
\mathcal{L}_{\rm soft}^{\rm MSSM} &=&-\left[
  \tQ^\dagger_i({\bf m^2_Q})_{ij}\tQ_j +\td^\dagger_{Ri}({\bf m^2_D})_{ij}\td_{Rj} +
  \tu^\dagger_{Ri}({\bf m^2_U})_{ij}\tu_{Rj}\right. \nonumber \\
&&\quad +\left.\tL^\dagger_i({\bf m^2_L})_{ij}\tL_j +
  \te^\dagger_{Ri}({\bf m^2_E})_{ij}\te_{Rj} +m_{H_u}^2 |H_u|^2
  +m_{H_d}^2 |H_d|^2 \right]  \nonumber \\
&&-\frac{1}{2}\left[M_1\bar{\lambda}_0\lambda_0 +
  M_2\bar{\lambda}_A \lambda_A+ M_3\bar{\tg}_B\tg_B\right]\nonumber\\
&&+\left[ ({\bf a_u})_{ij}\epsilon_{ab}\tQ_i^aH_u^b\tu_{Rj}^\dagger +
  ({\bf a_d})_{ij}\tQ_i^aH_{da}\td_{Rj}^\dagger
  +({\bf a_e})_{ij}\tL_i^a H_{da}\te_{Rj}^\dagger +{\rm h.c.}\right]  \nonumber \\
&&+\left[ bH_u^aH_{da} +{\rm h.c.}\right]\,.
\label{eq:mssmsoft}
\eea
The gaugino fields $\tg_B\, (B=1..8)$, $\lambda_A\, (A=1..3)$ and $\lambda_0$ transform
according to the adjoint representations of $SU(3)_c,\ SU(2)_L$ and $U(1)_Y$, respectively.

We use the RL convention for the fermion mass term,  
$\mathcal{L}_{mass} = -(\bar{\psi}_R^i m_{ij} \psi_L^j +{\rm h.c.)}$, in which
physical (real and diagonal) mass matrices read
\beq
\bm{m}_u=v_u {\bf V_{u_R}}{\bf f}_u^T{\bf V^\dagger_{u_L}}\,,\qquad
\bm{m}_d=v_d {\bf V_{d_R}}{\bf f}_d^T{\bf V^\dagger_{d_L}}\,,\qquad
\bm{m}_e=v_d {\bf V_{e_R}}{\bf f}_e^T{\bf V^\dagger_{e_L}}\,,
\label{eq:fermass}
\eeq
where $v_u,\ v_d$ are up- and down-higgs VEVs with $v=\sqrt{v^2_u + v^2_d} \simeq 174$~GeV.
The unitary rotation matrices ${\bf V_{\bullet}}$ transform 
gauge eigenstates (unprimed) to mass eigenstates (primed) as follows:
\begin{align}
u'_{Li}&=({\bf V_{u_L}})_{ij}\, u_{Lj}\, , & u'_{Ri}&=({\bf V_{u_R}})_{ij}\, u_{Rj}\, ,\nonumber \\
d'_{Li}&=({\bf V_{d_L}})_{ij}\, d_{Lj}\, , & d'_{Ri}&=({\bf V_{d_R}})_{ij}\, d_{Rj}\, ,\\
e'_{Li}&=({\bf V_{e_L}})_{ij}\, e_{Lj}\, , & e'_{Ri}&=({\bf V_{e_R}})_{ij}\, e_{Rj}\, .\nonumber
\label{eq:ferrot}
\end{align}
The Cabibbo-Kobayashi-Maskawa (CKM) matrix~\cite{ckm} is ${\bf V}_{CKM}={\bf V_{u_L}} {\bf V^\dagger_{d_L}}$. 
Due to different matrix diagonalization conventions, our rotation matrices ${\bf V_{\bullet}}$ are
hermitian conjugates of those in Ref.~\cite{andrew,Box:2008qs}.

We work in the Super-CKM (SCKM) basis~\cite{sckm} where gluino vertices remain flavor diagonal.
Here, the diagonalization of sfermion mass matrices proceeds in
two steps. First, the squarks and sleptons are rotated ``in parallel'' to their
fermionic superpartners
\begin{align}
\tu'_{Li}&=({\bf V_{u_L}})_{ij}\, \tu_{Lj}\, , & \tu'_{Ri}&=({\bf V_{u_R}})_{ij}\, \tu_{Rj}\, ,\nonumber \\
\td'_{Li}&=({\bf V_{d_L}})_{ij}\, \td_{Lj}\, , & \td'_{Ri}&=({\bf V_{d_R}})_{ij}\, \td_{Rj}\, ,\nonumber \\
\te'_{Li}&=({\bf V_{e_L}})_{ij}\, \te_{Lj}\, , & \te'_{Ri}&=({\bf V_{e_R}})_{ij}\, \te_{Rj}\, ,\nonumber \\
\tnu'_{Li}&=({\bf V_{e_L}})_{ij}\, \tnu_{Lj}\,,
\end{align}
where the SCKM scalar fields (primed) form supermultiplets with the corresponding fermion mass
eigenstates, {\it i.e.}, SCKM basis preserves the superfield structure after diagonalization of
fermions.
Next, $6 \times 6$ sfermion mass squared matrices in the SCKM basis are constructed:
\bea
\label{eq:mass2}
\bm{\mathcal{M}}^2_{\tu}&=&
 \begin{pmatrix}
 {\bf M}^2_{\tu LL}+\bm{m}^2_u+D(\tu_L)\mathds{1} & -{\bf M}^2_{\tu LR}+\mu \cot\beta \,\bm{m}_u \\
 -{\bf M}^{2\dagger}_{\tu LR}+\mu^* \cot\beta \,\bm{m}_u  & {\bf M}^2_{\tu RR}+\bm{m}^2_u+D(\tu_R)\mathds{1}
 \end{pmatrix} , \nonumber \\
\bm{\mathcal{M}}^2_{\td}&=&
 \begin{pmatrix}
 {\bf M}^2_{\td LL}+\bm{m}^2_d+D(\td_L)\mathds{1} & -{\bf M}^2_{\td LR}+\mu \tan\beta \,\bm{m}_d \\
 -{\bf M}^{2\dagger}_{\td LR}+\mu^* \tan\beta \,\bm{m}_d  & {\bf M}^2_{\td RR}+\bm{m}^2_d+D(\td_R)\mathds{1}
 \end{pmatrix} ,\\
\bm{\mathcal{M}}^2_{\te}&=&
 \begin{pmatrix}
 {\bf M}^2_{\te LL}+\bm{m}^2_e+D(\te_L)\mathds{1} & -{\bf M}^2_{\te LR}+\mu \tan\beta \,\bm{m}_e \\
 -{\bf M}^{2\dagger}_{\te LR}+\mu^* \tan\beta \,\bm{m}_e  & {\bf M}^2_{\te RR}+\bm{m}^2_e+D(\td_R)\mathds{1}
 \end{pmatrix} , \nonumber 
\eea
where $D(\tilde f)$ are the hypercharge $D-$term contributions to the corresponding
sfermions, $\mathds{1}$ is the $3\times 3$ unit matrix, 
$\bm{m}_f$ are the diagonal fermion mass matrices of Eq.~(\ref{eq:fermass}), 
and the flavor-changing entries are contained in rotated SSB matrices,
\beq
\begin{array}{lcccr}
{\bf M}^2_{\tu LL}={\bf V_{u_L} m^2_Q V^\dagger_{u_L}}\,, &\ &
{\bf M}^2_{\tu RR}={\bf V_{u_R} m^2_U V^\dagger_{u_R}}\,, &\ &
{\bf M}^2_{\tu LR}={\bf V_{u_L} a^*_u V^\dagger_{u_R}}\,, \\
{\bf M}^2_{\td LL}={\bf V_{d_L} m^2_Q V^\dagger_{d_L}}\,, &\ &
{\bf M}^2_{\td RR}={\bf V_{d_R} m^2_D V^\dagger_{d_R}}\,, &\ &
{\bf M}^2_{\td LR}={\bf V_{d_L} a^*_d V^\dagger_{d_R}}\,, \\
{\bf M}^2_{\te LL}={\bf V_{e_L} m^2_L V^\dagger_{e_L}}\,, &\ &
{\bf M}^2_{\te RR}={\bf V_{e_R} m^2_E V^\dagger_{e_R}}\,, &\ &
{\bf M}^2_{\te LR}={\bf V_{e_L} a^*_e V^\dagger_{e_R}}\,.
\end{array}
\label{eq:sckm}
\eeq
Note that the squark doublet mass-squared SSB matrix ${\bf m^2_Q}$ is rotated differently for
$\bm{\mathcal{M}}^2_{\tu}$ and $\bm{\mathcal{M}}^2_{\td}$.
Due to the absence of right-handed neutrino states in the MSSM, the sneutrino mass-squared matrix is a 
$3\times 3$ matrix of the form,
\bea
\bm{\mathcal{M}}^2_{\tnu} &=& {\bf V_{e_L} m^2_L V^\dagger_{e_L}} +D(\tnu_L)\mathds{1}\,.
\label{eq:nu_mass2}
\eea
Finally, the mass squared matrices (\ref{eq:mass2}) and (\ref{eq:nu_mass2}) are diagonalized to obtain sfermion mass
eigenstates. These mass eigenstates are labelled in ascending mass order.

To incorporate neutrino masses, we employ type-I seesaw mechanism where the MSSM is extended by three gauge singlet 
superfields $\hat{N}^c_i$ each of whose
fermionic component is the left-handed anti-neutrino and  
scalar component is $\tnu^\dagger_{Ri}$. The extended superpotential has the form shown in
Eq.~(\ref{eq:RHNspot}). 
The neutrino Yukawa and Majorana mass matrices are diagonalized in analogy with Eq.~(\ref{eq:fermass}) by unitary
matrices ${\bf V_{\nu_L}},{\bf V_{\nu_R}}$ and ${\bf V_N}$ according to
\beq
\bm{m}_D=v_u {\bf V_{\nu_R}}{\bf f}_{\nu}^T{\bf V^\dagger_{\nu_L}},\qquad
{\bf V^{\phantom{T}}_N}{\bf M^{\phantom{T}}_N}{\bf V}^T_{\bf N}={\rm diag}(M_{N_1},M_{N_2},M_{N_3}),
\label{eq:nu_diag}
\eeq
where $\bm{m}_D$ is the Dirac neutrino mass matrix. 
Apriori the eigenvalues $M_{N_i}$ are arbitrary, but the observed large difference
between neutrino and charged lepton masses strongly suggests that the scale of Majorana masses
$M_{Maj}\equiv {\rm max}(M_{N_3})$ is much higher than the weak scale. 
Additional soft SUSY breaking terms should also be included so that the Lagrangian becomes,
\beq
\mathcal{L}_{soft} = \mathcal{L}_{soft}^{MSSM} - \tnu^{\dagger}_{Ri} (\mns)_{ij} \tnu_{Rj}
  +\left[ (\banu)_{ij}\epsilon_{ab} \tL_i^a \tH_u^b \tnu^{\dagger}_{Rj}
         +\frac{1}{2}(\bbnu)_{ij}\tnu_{Ri}\tnu_{Rj} + {\rm h.c.} \right] \,.
\label{eq:ssb}
\eeq
The eigenvalues of matrices $\banu$ and $\bbnu$ and  the square roots of the eigenvalues of $\mns$, are
 assumed to be of the order of the weak scale $M_{weak}$.


Above the scale $M_{Maj}$, the mass matrix for left-handed neutrinos is given by Eq.~(\ref{eq:seesaw}). 
This is the minimal or type-I seesaw mechanism~\cite{seesaw1}. 
At low energies all RHNs decouple and the theory is governed by the effective superpotential,
\beq
\hat{f}_{eff}=\hat{f}_{MSSM}+
\frac{1}{2} \kappa_{ij} \epsilon_{ab}\hat{L}^a_i\hat{H}_u^b \epsilon_{df}\hat{L}^d_j\hat{H}_u^f \, .
\label{eq:effspot}
\eeq
Here, $\kappa$ is a $3 \times 3$ complex symmetric coupling matrix that breaks lepton number explicitly, and is determined by
matching conditions at RHN thresholds,
\beq
\left.(\kappa)_{ij} \right\vert _{M^-_{N_k}} = \left.(\kappa)_{ij} \right\vert _{M^+_{N_k}} + \left. (\bfn)_{ik} \frac{1}{M_{N_k}} (\bfn^T)_{ik} \right\vert _{M^+_{N_k}}
\label{eq:Kmatch}
\eeq
where ${M^+_{N_k}}({M^-_{N_k}})$ denotes the value as the scale of decoupling of
the k-th generation RHN $M_{N_k}$ is approached from above (below).

After electroweak symmetry breaking, the Majorana mass matrix for light left-handed neutrinos is simply
$\mathcal{M}_{\nu}=-\kappa v^2_u$.
The symmetric $3 \times 3$  matrix $\mathcal{M}_{\nu}$ can be diagonalized as
\beq
  {\bf U_{\nu}}\mathcal{M}_{\nu}{\bf U}^T_{\bf \nu} = \bm{m}_\nu\,,
\eeq
where ${\bf U_{\nu}}$ is a unitary matrix and $\bm{m}_\nu$ is a diagonal matrix of the physical
neutrino masses $m_1,\ m_2$ and $m_3$. 
In labeling the (real non-negative) mass eigenstates we follow the usual convention 
that 1 and 2 denote states with the smallest mass-squared difference and $m_1 < m_2$.
The Maki-Nakagava-Sakata (MNS) mixing matrix of physical neutrinos~\cite{mns} then has the form 
${\bf V}_{MNS}={\bf V_{e_L}} {\bf U_{\nu}^\dagger}$. Note that as a consequence of the seesaw mechanism, 
 ${\bf U_{\nu}}$ is in general
different from the neutrino Yukawa diagonalization matrix ${\bf V_{\nu_L}}$ defined in Eq.(\ref{eq:nu_diag}).

The MNS matrix can be parametrized as
\beq
\label{eq:mns}
{\bf V}_{MNS}=
 \begin{pmatrix}
 c_{12}c_{13} & s_{12}c_{13} & s_{13}e^{-i\delta} \\
 -c_{23}s_{12}-s_{23}s_{13}c_{12}e^{i\delta} & 
  c_{23}c_{12}-s_{23}s_{13}s_{12}e^{i\delta} & s_{23}c_{13}\\
 s_{23}s_{12}-c_{23}s_{13}c_{12}e^{i\delta} &
  -s_{23}c_{12}-c_{23}s_{13}s_{12}e^{i\delta} & c_{23}c_{13}
 \end{pmatrix}
 \times {\rm diag}(e^{i\frac{\phi_1}{2}},e^{i\frac{\phi_2}{2}},1)\,,
\eeq
with $s_{ij}=\sin\theta_{ij}$ and $c_{ij}=\cos\theta_{ij}$, $\theta_{ij}$ are mixing angles and
the Dirac and Majorana CP-violating phases are $\delta,\ \phi_{1,2} \in [0,2\pi]$, respectively. 

The addition of the $\hat{N}^c_i$ superfields also leads to an expansion of the sneutrino mass-squared matrix:
it is now a $12 \times 12$ matrix, so that the relevant part of Lagrangian is
\bea
\label{eq:snumass}
\mathcal{L} & \ni & -\frac{1}{2} \tilde{n}^\dagger
 \begin{pmatrix}
\bcm2_{L^\dagger L} & \bm{0} & \bcm2_{L^\dagger R} & v_u \bfn^* {\bf M}^T_{\bf N}  \\
\bm{0} & (\bcm2_{L^\dagger L})^T & v_u \bfn {\bf M}^\dagger_{\bf N}  & (\bcm2_{L^\dagger R})^* \\
(\bcm2_{L^\dagger R})^\dagger & v_u {\bf M^{\phantom{\dagger}}_N}\bfn^\dagger & \bcm2_{R^\dagger R} & -\bbnu^\dagger \\
v_u {\bf M^*_N} \bfn^T & (\bcm2_{L^\dagger R})^T & -\bbnu & (\bcm2_{R^\dagger R})^T 
 \end{pmatrix}
\tilde{n}\,,
\eea
where $\tilde{n}^T \equiv (\tnu_L^T,\tnu^{\dagger}_L,\tnu_R^T,\tnu^{\dagger}_R)$, $\bm{0}$ is the 
$3\times 3$ null matrix and
\bea
 \bcm2_{L^\dagger L} &=& {\bf m^2_L} + v_u^2 \bfn^* \bfn^T + D(\tnu_L)\mathds{1}\,, \nonumber \\
 \bcm2_{R^\dagger R} &=& \mns + v_u^2 \bfn^T \bfn^* + {\bf M^{\phantom{\dagger}}_N}{\bf M}^\dagger_{\bf N}\,, \\
 \bcm2_{L^\dagger R} &=& - v_u \banu^* + \mu v_d \bfn^*\,. \nonumber
\eea
From this structure we see that there is sneutrino-antisneutrino mixing for right-handed states
introduced by $\bbnu$ with no corresponding terms in the left-handed sector.
The Majorana mass matrix ${\bf M_N}$ contributes to the mass of the right-handed states and also 
results in the mixing of right-handed anti-sneutrino states with left-handed sneutrinos. 
Since ${\bf M_N}$ eigenvalues are much larger than the rest of the SSB parameters the
matrix exhibits a seesaw type behavior, similar to the one for neutrinos: 
the $6\times 6$ L-L block is of $\mathcal{O}(M_{weak}^2)$, while R-L blocks are of
$\mathcal{O}(M_{weak}M_{Maj})$ and the R-R block is $\mathcal{O}(M_{Maj}^2)$.
Therefore the right-handed sneutrinos decouple and the phenomenologically relevant left-handed
sneutrinos have a mass-squared matrix of the familiar MSSM form (\ref{eq:nu_mass2}).

\section{Description of ISAJET-M}
\label{app:code}

\TABLE{
\begin{tabular}{|l|c||l|c|}
\hline
parameter & value & parameter & value \\
\hline
$M_Z$ & 91.1876                      & 	$m_t$ &  171\\
$1 / \alpha^{\MSb}(M_Z)$ & 127.918   &  $m_b - m_c$ & 3.42\\
$\alpha_s^{\MSb}(M_Z)$ & 0.1176      &	$m_e \times 10^3$ &  0.511\\
$\sin^2 \theta_W (M_Z)$ & 0.23122    &  $m_{\mu}$ &  0.10566\\
$m_u(2 \,\text{GeV})$ & 0.003 		     &  $m_{\tau}$ & 1.77699 \\
$m_d(2 \,\text{GeV})$ & 0.006 		     &  $\sin \theta^{CKM}_{12}$ & 0.22715\\
$m_s(2 \,\text{GeV})$ & 0.095 		     &  $\sin \theta^{CKM}_{23}$ & 0.04161\\
$m_c(m_c)^{\MSb}$ & 1.25 	     &  $\sin \theta^{CKM}_{13}$ & 0.003682\\
$m_b(m_b)^{\MSb}$ & 4.20 	     &  $\delta^{CKM}$ & 1.0884\\
\hline
\end{tabular}
\caption{SM input parameters~\cite{pdg} for ISAJET-M. All masses are in GeV and $\delta^{CKM}$ is in radians.}
\label{tab:sm}}
%

In this section we describe the algorithm used to perform the calculation. Our code, named
ISAJET-M, is ISAJET 7.78~\cite{isajet} modified to evolve all couplings and SSB parameters, including
the neutrino sector, in full matrix form at the 2-loop level.
The following Standard Model parameters are used as inputs: fermion masses, the
Z-boson pole mass $M_Z$, the fine structure constant $\alpha^{\MSb}(M_Z)$, the
strong coupling constant $\alpha_s^{\MSb}(M_Z)$ and CKM angles in the ``standard parametrization''; see Table~\ref{tab:sm}.

In the first step $\alpha^{\MSb}$ and $\alpha_s^{\MSb}$ are evolved from $Q=M_Z$ down
to $Q=2$~GeV using 2-loop QCD$\times$QED RGEs with the additional 3rd QCD
loop~\cite{Arason,qcd3l}. We use step-function decoupling of fermions at the scale of their
running mass and include finite threshold corrections at 2 loops according to Ref.~\cite{Chetyrkin}. 
Then we compute the running lepton masses $m_l^{\MSb}(2 \, \rm{GeV})$ from their pole masses
$m_l$ using the 1-loop expression in the $\MSb$ scheme~\cite{Arason}:
\beq
  m_l^{\MSb}(Q)= m_l \left[1-\frac{\alpha^{\MSb}(Q)}{\pi}\left(1+\frac{3}{4} \ln \frac{Q^2}{m_l^2}\right)\right]\,.
\eeq
Next, the two gauge couplings and all SM fermion masses (except for the top
mass) are run up to $M_Z$. Here fermion masses are converted from $\MSb$ to
$\DRb$ scheme using formulae given in Ref.~\cite{Avdeev,Martin}:
\bea
 m_b^{\DRb}(M_Z) 
 &=& m_b^{\MSb}(M_Z)\left(1-\frac{\alpha_s}{3\pi}-\frac{29\alpha_s^2}{72\pi^2}
     +\frac{3g_2^2}{128\pi^2}+\frac{13g_1^2}{1920\pi^2}\right)_{\MSb}\,,  \nonumber \\
 m_c^{\DRb}(M_Z) 
 &=& m_c^{\MSb}(M_Z)\left(1-\frac{\alpha_s}{3\pi}-\frac{29\alpha_s^2}{72\pi^2}
     +\frac{3g_2^2}{128\pi^2}+\frac{g_1^2}{1920\pi^2}\right)_{\MSb}\,, \\
 m_{\tau}^{\DRb}(M_Z) 
 &=& m_{\tau}^{\MSb}(M_Z)
     \left(1+\frac{3g_2^2}{128\pi^2}-\frac{9g_1^2}{640\pi^2}\right)_{\MSb}\,.  \nonumber
\eea
For the lighter SM fermions these conversion corrections are neglected, {\it i.e.}, we take
$m_f^{\DRb}=m_f^{\MSb}$ at $M_Z$ for $f=u,\ d,\ s,\ e,\ \mu$.

For the top quark, we obtain the running mass at $Q=m_t$ using the 2-loop QCD expression~\cite{Bednyakov:2002sf},
\beq
  m_t^{\DRb}(m_t)= m_t
  \left[1+\frac{5}{3}\frac{\alpha_s(m_t)}{\pi}+\left(\frac{\alpha_s(m_t)}{\pi}\right)^2
  \Sigma_t^{2loop} \right]\,,
\eeq
where $\Sigma_t^{2loop}$ is the 2-loop piece.
Note that we activate the top quark at the scale of its mass, so we have a 5-flavor scheme
below $Q=m_t$ and all 6 flavors above it.

The obtained $\DRb$ values of SM fermion masses are then substituted into Eq.~(\ref{eq:fermass}) to
calculate the Yukawa matrices at $M_Z$ in the gauge eigenbasis. 
A choice of basis is made by specifying fermion rotation matrices ${\bf V_{\bullet}}$ defined in Eq.~(\ref{eq:fermass}).
ISAJET-M also has options for performing calculations 
in the unmixed and dominant third-family approximations.

The obtained weak scale values of gauge couplings and Yukawa matrices are evolved to the Grand Unification
scale $M_{GUT}$ via MSSM RGEs in the $\DRb$ scheme. 
The GUT scale $M_{GUT}$ is defined to be the scale at which $g_1=g_2$, 
where $g_1=\sqrt{5/3} \, g'$ is the hypercharge coupling in the GUT-scale normalization. We do
not impose an exact unification of the strong coupling ($g_3=g_1=g_2$) at $M_{GUT}$,
assuming that the resulting few percent
discrepancy comes from GUT-scale threshold corrections~\cite{yamada}.

For the evolution of gauge and Yukawa couplings, we use a multiscale effective theory approach proposed
in Ref.~\cite{Castano}, where heavy degrees of freedom are integrated out at each particle
threshold. 
In Appendix~\ref{app:rges}, we list the corrected Yukawa 1-loop RGEs from an unpublished erratum of Ref.~\cite{Castano}.
In the second-loop RGE terms we change from MSSM formulae~\cite{rge} 
to SM expressions~\cite{Barger:1992ac,Box:2008qs} at a single scale, 
$Q=M_{SUSY}\equiv \sqrt{m_{\ttop_L}m_{\ttop_R}}$, thus 
introducing an error of 3-loop order which is negligible. 

This ``step beta-function approach'' produces continuous matching conditions across thresholds. However,
decoupling of a heavy particle also introduces finite shifts in RGE parameters~\cite{shifts}; a
similar effect has long been known in QCD, where the decoupling of heavy quarks leads to shifts
in the running masses of the light quarks~\cite{Wetzel:1981qg}.
Expressions for shifts induced by decoupling of each individual sparticle depend on the ordering of sparticle
spectrum and are not yet known for the general case.
Therefore, we implement these sparticle-induced finite shifts (to all three generations) 
collectively at a common scale $Q=M_{SUSY}$ in the basis where Yukawa matrices are diagonal; 
we use 1-loop expressions of Ref.~\cite{Pierce} without logarithmic terms that have already been resummed
by the RGE evolution. For the top Yukawa coupling additional 2-loop SUSY-QCD corrections are
included according to Ref.~\cite{Bednyakov:2002sf}. These finite threshold corrections are particularly
important for GUT theories since they change ratios of Yukawa couplings from those at the weak
scale, as was emphasized in Ref.~\cite{Ross:2007az}.
This multiscale approach is a generalization of the one used by the standard ISAJET, as is described in 
detail in Ref.~\cite{Baer:2005pv},
and is preferred to single-scale decoupling when the sparticle mass hierarchy is large 
(as appears, for example, in the HB/FP region~\cite{dm:fp} of mSUGRA).

At $M_{GUT}$, the SSB boundary
conditions are imposed and all SSB parameters along with gauge and
Yukawa couplings are evolved back down to the weak scale $M_{Z}$. 
For the SSB parameters we use 2-loop RGEs from Ref.~\cite{rge} with the following conversion
between notations: ${\bf f_\bullet} \equiv {\bf Y_\bullet}^T$, ${\bf a_\bullet} \equiv -{\bf h_\bullet}^T$, $b\equiv -B$.
Unlike the gauge and Yukawa couplings, where
beta-functions change at every threshold, the SSB beta-functions remain those of the MSSM all the way down
to $M_Z$. 
We do not take into account threshold effects from the appearance of new couplings in the region of
broken supersymmetry introduced in Ref.~\cite{andrew,Box:2008qs}. 
The entire parameter set is iteratively run between $M_{Z}$ and
$M_{GUT}$ using full 2-loop RGEs in matrix form until a stable solution is obtained.
After each iteration the Higgs potential is minimized and sparticle/Higgs masses are recalculated. The obtained mass
spectrum is used in the next iteration to appropriately account for sparticle threshold effects on
the RGE evolution.

The minimization of the RGE-improved potential is done using the tadpole
method~\cite{tadpole}. We include 1-loop contributions from third generation sfermions, which 
dominate, as well as contributions from charginos, neutralinos and Higgs bosons.
Computation is done at an optimized scale $Q=M_{SUSY}$, which effectively accounts for the leading 2-loop
corrections.
Since tadpoles that contribute to $\mu$ and $b$ strongly depend on
the parameters themselves, an iterative procedure is employed. Calculations of
$\mu$, $b$ and tadpoles are iterated until consistent values with a precision of $0.1\%$ are
obtained; usually this requires 3-4 iterations.

In the computation of sparticle masses we use SSB parameters extracted at their respective mass scales. Then,
the SSB matrices are assembled and rotated to the SCKM basis as in Eq.~(\ref{eq:sckm}). 
The resultant matrices are plugged into the sfermion mass-squared matrices (\ref{eq:mass2}). Instead of
diagonalizing the full $6\times 6$ matrices (\ref{eq:mass2}) we diagonalize three $2\times 2$ submatrices, thereby
neglecting intragenerational mixings, which are required to be small by experimental limits on
flavor changing neutral current processes~\cite{Ciuchini:2007ha}.
For the finite corrections, the full expressions of Ref.~\cite{Pierce} for 1-loop self-energies are used.

For the SUSY-seesaw  we also use the multiscale approach that is mandatory to obtain correct
values in the neutrino sector in the case of hierarchical RHNs~\cite{Antusch:2002rr}.
Our approach is identical to the one used in the REAP code~\cite{Antusch:2005gp}. 
Above the scale of the heaviest RHN we have a full MSSM+RHN setup, with
the MSSM RGEs extended to include full 2-loop equations for $\bfn$, $\banu$, ${\bf m^2_{\tnu_R}}$
and ${\bf M}_N$~\cite{casas,Ibarra:2008uv}. 
As mentioned earlier, bilinear coupling matrix ${\bf b}_\nu$ only introduces mixings for right-handed sneutrino states whose masses
are $\mathcal{O}(M_{Maj})$ and is thus irrelevant for our analysis. 
We also take into account additional contributions to RGEs of ordinary MSSM parameters due to 
the RHN superfields up to 2-loop order~\cite{casas,Ibarra:2008uv,Antusch:2005gp,Casas:2000pa}. 
For large neutrino Yukawa couplings these additional RGE terms 
cause changes to the MSSM sparticle spectrum that can have significant consequences for experimental rates~\cite{Baer:2000hx,msugrarhn}.
Below the scale of the lightest RHN, the RGEs are those of the MSSM, plus additional equations for the
coupling $\kappa$ of the dimension-5 effective neutrino operator (\ref{eq:effspot})
that are included at 2-loop level~\cite{Antusch:2005gp}.
In the intermediate region, both $\kappa$ and some elements of $\bfn$ and ${\bf M}_N$ are present. 
Transitions between them are done using matching conditions at RHN thresholds Eq.~(\ref{eq:Kmatch}).
We match only at tree-level and neglect small finite threshold corrections.
The position of RHN thresholds $M_{N_k}$ are determined by the eigenvalues of the RGE-evolving
Majorana mass matrix ${\bf M}_N$ at those scales. 
Note that Eq.~(\ref{eq:Kmatch}) is valid only in the basis where ${\bf M}_N$ is diagonal at the threshold,
which is different from the original basis at $M_{GUT}$ due to RGE effects. 
Below the scale $Q=M_{N_k}$, the $k-$th RHN superfields are absent from the theory so we remove the
corresponding columns of $\bfn$ and $\banu$ matrices and the $k-$th row and column of ${\bf m^2_{\tnu_R}}$ and ${\bf M}_N$
in the basis where ${\bf M}_N$ is diagonal.

Since some of the RGE parameters (gauge and Yukawa couplings) are defined at the weak scale, while the rest are
set at $M_{GUT}$, an iterative procedure is employed to solve the RGEs. 
Once a stable solution to the RGEs is obtained, the decay width and branching ratios for all the sparticles and
Higgs bosons are calculated. This step as well as the previous computation of the mass spectrum is
performed using standard ISAJET subroutines~\cite{isajet}. 
A graphical outline of our code, for a SUSY-seesaw model that uses top-down approach for the neutrino sector
({\it i.e.}, $\bfn$, ${\bf M}_N$ are inputted at the GUT scale and physical light neutrino masses and mixing
parameters are derived results), is shown in Fig.~\ref{fig:evolchart}.

ISAJET-M has facility to calculate and return
the neutralino relic density $\Omega_{\tz_1} h^2$, neutralino-nucleon elastic cross sections,
branching fractions for $b\rightarrow s\gamma$ and $B_s \rightarrow \mu\mu/\tau\tau$ decays,
and supersymmetric contributions to the muon anomalous magnetic moment $\Delta a_\mu \equiv (g-2)_\mu /2$.
These computations are done, respectively, by the IsaReD~\cite{isared}, IsaReS~\cite{isares}, 
IsaBSG~\cite{isabsg}, IsaBMM~\cite{isabmm} and IsaAMU~\cite{isaamu} codes from the IsaTools package. 
In addition, for scenarios with massive neutrinos, branching fractions for 
lepton flavor violating decays $l_i\rightarrow l_j \gamma$ and $l_i \rightarrow 3l_j$ 
as well as rates for $\mu \rightarrow e$
conversion in several nuclei can be calculated using expressions from Ref.~\cite{hisano}.

\section{Yukawa RGEs}
\label{app:rges}

The following are the 1-loop RGEs for MSSM Yukawa coupling matrices with sparticle/higgs thresholds from
Ref.~\cite{Castano} with corrections from their unpublished erratum implemented: 

\bea
(4\pi)^2 \frac{d(\byu)_{ij}}{dt}
&=& \frac{3}{2}\left(s^2 \theta_h+c^2 \theta_H\right)(\byu \bfu^* \byu)_{ij} 
+\frac{1}{2}\left(c^2 \theta_{\th}+s^2 \theta_{\tH}\right)
    \sum_{k=1}^{N_{\td}}(\byd)_{ik}(\bfd^*)_{kl}(\byu)_{lj} \nonumber \\
&& +\frac{1}{2}\left(s^2 \theta_{\th}+c^2 \theta_{\tH}\right)
    \left[2(\byu)_{il} \sum_{k=1}^{N_{\tQ}}(\bfu^*)_{lk}(\byu)_{kj}
          +\sum_{k=1}^{N_{\tu}}(\byu)_{ik}(\bfu^*)_{kl}(\byu)_{lj}\right] \nonumber \\
&& +\frac{1}{2}\left(c^2\theta_h + s^2\theta_H -4c^2(\theta_h-\theta_H)\right)(\byd \bfd^* \byu)_{ij} \nonumber \\
&& +(\byu)_{ij} 
    \left[(s^2\theta_h +c^2) \Tr \lbrace 3\bfu^* \byu \rbrace 
          +c^2(\theta_h-1) \Tr \lbrace 3\bfd^* \byd +\bfe^* \bye \rbrace \right] \nonumber \\
&& -(\byu)_{ij}
    \left[ \frac{3}{5}g_1^2 
          \left\{ \frac{17}{12}+\frac{3}{4}\theta_{\th}
	         -\left( \frac{1}{36}\theta_{\tQ_j}+\frac{4}{9}\theta_{\tu_i}
		        + \frac{1}{4}\theta_{\th} \right) \theta_{\tB} \right\} \right. \nonumber \\
&& \left. \qquad \ \quad \;
         +g_2^2 \left\{ \frac{9}{4}+\frac{9}{4}\theta_{\th}
	         -\frac{3}{4}\left( \theta_{\tQ_j}+\theta_{\th}\right) \theta_{\tW} \right\} 
         +g_3^2 \left\{ 8-\frac{4}{3}\left( \theta_{\tQ_j}+\theta_{\tu_i}\right) \theta_{\tg} \right\} \right]\,, \nonumber \\
(4\pi)^2 \frac{d(\byd)_{ij}}{dt}
&=& \frac{3}{2}\left(c^2 \theta_h+s^2 \theta_H\right)(\byd \bfd^* \byd)_{ij} 
+\frac{1}{2}\left(s^2 \theta_{\th}+c^2 \theta_{\tH}\right)
    \sum_{k=1}^{N_{\tu}}(\byu)_{ik}(\bfu^*)_{kl}(\byd)_{lj} \nonumber \\
&& +\frac{1}{2}\left(c^2 \theta_{\th}+s^2 \theta_{\tH}\right)
    \left[2(\byd)_{il} \sum_{k=1}^{N_{\tQ}}(\bfd^*)_{lk}(\byd)_{kj}
          +\sum_{k=1}^{N_{\td}}(\byd)_{ik}(\bfd^*)_{kl}(\byd)_{lj}\right] \nonumber \\
&& +\frac{1}{2}\left(s^2\theta_h + c^2\theta_H -4s^2(\theta_h-\theta_H)\right)(\byu \bfu^* \byd)_{ij} \nonumber \\
&& +(\byd)_{ij} 
    \left[s^2(\theta_h-1)\Tr \lbrace 3\bfu^* \byu \rbrace 
          +(c^2\theta_h+s^2) \Tr \lbrace 3\bfd^* \byd +\bfe^* \bye \rbrace \right] \nonumber \\
&& -(\byd)_{ij}
    \left[ \frac{3}{5}g_1^2 
          \left\{ \frac{5}{12}+\frac{3}{4}\theta_{\th}
	         -\left( \frac{1}{36}\theta_{\tQ_j}+\frac{1}{9}\theta_{\td_i}
		        +\frac{1}{4}\theta_{\th} \right) \theta_{\tB} \right\} \right. \nonumber \\
&& \left.\qquad \ \quad \;
         +g_2^2 \left\{ \frac{9}{4}+\frac{9}{4}\theta_{\th}
	         -\frac{3}{4}\left( \theta_{\tQ_j}+\theta_{\th}\right) \theta_{\tW} \right\} 
         +g_3^2 \left\{ 8-\frac{4}{3}\left( \theta_{\tQ_j}+\theta_{\td_i}\right) \theta_{\tg} \right\} \right]  \,, \nonumber \\
(4\pi)^2 \frac{d(\bye)_{ij}}{dt}
&=& \frac{3}{2}\left(c^2 \theta_h+s^2 \theta_H\right)(\bye \bfe^* \bye )_{ij} \nonumber \\
&& +\frac{1}{2}\left(c^2 \theta_{\th}+s^2 \theta_{\tH}\right)
    \left[2(\bye)_{il} \sum_{k=1}^{N_{\tL}}(\bfe^*)_{lk}(\bye)_{kj}
          +\sum_{k=1}^{N_{\te}}(\bye)_{ik}(\bfe^*)_{kl}(\bye)_{lj}\right] \nonumber \\
&& +(\bye)_{ij} 
    \left[s^2(\theta_h-1)  \Tr \lbrace 3\bfu^* \byu \rbrace 
          +(c^2\theta_h+s^2) \Tr \lbrace 3\bfd^* \byd +\bfe^* \bye \rbrace \right] \nonumber \\
&& -(\bye)_{ij}
    \left[ \frac{3}{5}g_1^2 
          \left\{ \frac{15}{4}+\frac{3}{4}\theta_{\th}
	         -\left( \frac{1}{4}\theta_{\tL_j}+\theta_{\te_i}
		        +\frac{1}{4}\theta_{\th} \right) \theta_{\tB} \right\} \right. \nonumber \\
&& \left.\qquad \ \quad \;
         +g_2^2 \left\{ \frac{9}{4}+\frac{9}{4}\theta_{\th}
	               -\frac{3}{4}\left( \theta_{\tL_j}+\theta_{\th}\right) \theta_{\tW} \right\}  \nonumber \right]\,, 
\eea
where $s=\sin \alpha,\ c=\cos \alpha$, $\alpha$ is Higgs mixing angle 
and the various $\theta_{\mathcal{P}}$'s are equal to zero below the mass threshold of the respective particle and equal to
one above it. The contributions from neutrino Yukawa couplings can be found
in Refs.~\cite{casas,Ibarra:2008uv}.


\end{document}